\RequirePackage[l2tabu,orthodox]{nag}

\documentclass[12pt,a4paper,onecolumn,oneside,notitlepage]{article}
\usepackage[left=3.25cm,right=3.25cm,top=4cm,bottom=4cm]{geometry}
\usepackage[onehalfspacing]{setspace}

\usepackage{natbib}
\usepackage[english]{babel}
\usepackage[T1]{fontenc}
\usepackage[utf8]{inputenc}
\usepackage{hyphenat}
\usepackage{microtype}
\usepackage{verbatim}
\usepackage{dsfont}

\usepackage{amsmath}
\usepackage{amsfonts}
\usepackage{amssymb}
\usepackage{amsthm}
\usepackage{nicefrac}
\usepackage{mathtools}
\usepackage{bbm} 

\usepackage{float}
\usepackage{makeidx}
\usepackage{tabularx}
\usepackage{graphicx}
\usepackage{epstopdf}
\usepackage{booktabs}
\usepackage{pdflscape}
\usepackage{longtable}
\usepackage{seqsplit}
\usepackage{rotating}
\usepackage{adjustbox}
\usepackage{multirow}
\usepackage{multicol}
\usepackage{dcolumn}
\usepackage[section]{placeins} \usepackage{makecell}

\usepackage{soul}
\usepackage{blindtext}
\usepackage{todonotes}
\usepackage{appendix}
\usepackage{titlesec}
\usepackage{enumerate}
\usepackage{thmtools}
\usepackage{tcolorbox}
\usepackage{colortbl} \usepackage[nopar]{lipsum}
\usepackage{textcomp}
\usepackage{xcolor}
\definecolor{darkRed}{RGB}{144,0,0}
\definecolor{darkBlue}{RGB}{0,0,144}
\definecolor{darkGreen}{RGB}{0,144,0}
\definecolor{darkgray}{rgb}{0.7, 0.7, 0.7}
\definecolor{lightgray}{rgb}{0.9, 0.9, 0.9}

\usepackage{hyperref}
\hypersetup{
    colorlinks=true,
    breaklinks=true,
    bookmarksnumbered=true,
    bookmarksopen=true,
    bookmarksdepth=2,
    citecolor=darkRed,
    linkcolor=darkRed,
    urlcolor=darkGreen,
    pdftitle={occ2vec},
    pdfauthor={Nicolaj Søndergaard Mühlbach}
}

\titleformat{\section}[block]{\large\bfseries\centering}{\thesection.}{0.5em}{}
\titleformat{\subsection}[block]{\normalsize\itshape}{\thesubsection.}{0.5em}{}
\titleformat{\subsubsection}[block]{\normalsize\itshape\centering}{\thesubsubsection.}{0.5em}{}

\renewcommand{\thesection}{\Roman{section}}
\renewcommand{\thesubsection}{\Alph{subsection}}
\renewcommand{\thesubsubsection}{\Alph{subsection}.\arabic{subsubsection}}

\theoremstyle{plain}

\newtheorem{definition}{Definition}

\usepackage[linesnumbered,ruled,vlined]{algorithm2e}

\SetKwInput{KwInput}{Input}                \SetKwInput{KwOutput}{Output}

\graphicspath{{../../figures/}{../../figures/validation/}{../../figures/assessment/}{../../figures/comparison/}{../../figures/misc/}{figures/}{figures/validation/}{figures/assessment/}{figures/comparison/}{figures/misc/}}

\makeatletter
\providecommand*{\input@path}{}
\g@addto@macro\input@path{{../../tables/}{../../tables/misc/}{../../tables/validation/}{../../tables/assessment/}{../../tables/comparison/}{tables/}{tables/misc/}{tables/validation/}{tables/assessment/}{tables/comparison/}}\makeatother

\newcommand{\DimRed}{tsne} \newcommand{\OccToVec}{\textbf{occ2vec}} 
\newcommand{\OccToVecS}{\OccToVec\SpaceBar}
\newcommand{\EducVar}{mean_education}
\newcommand{\textquote}[1]{``#1''}

\newcommand{\VerticalSpace}{\vspace{0in}} \newcommand{\VerticalSpaceFloat}{\vspace{0.166667in}} 

\newcommand{\SingleFigMaxHeight}{0.3}

\newcommand{\SpaceBar}{\hspace{0.2916667em}}

\usepackage{enumitem} 
\setlist[description]{leftmargin=\parindent,labelindent=\parindent}

\usepackage{caption}
\usepackage{subcaption}
\newcommand{\RomNum}[1]{\uppercase\expandafter{\romannumeral #1\relax}}

\long\def\symbolfootnote[#1]#2{\begingroup\def\thefootnote{\fnsymbol{footnote}}\footnote[#1]{#2}\endgroup}

\begin{document}

\begin{titlepage}
\pagenumbering{Alph}
	\newcommand{\mytitle}{\Large{\textbf{occ2vec: A principal approach to representing occupations using natural language processing}}}

    \thispagestyle{empty}
    \setlength{\parindent}{0cm}
    \renewcommand{\thefootnote}{\fnsymbol{footnote}}
    
    \begin{center}

\mytitle\symbolfootnote[1]
        {
        \noindent
The author is grateful to Susan Athey, Victor Chernozhukov, Daron Acemoglu, Andrew Scott, Elisa Macchi, Erik Christian Montes Shütte, Mathias Siggard, and Thyge Enggaard,  for helpful comments and suggestions, and to the Department of Economics, Massachusetts Institute of Technology (MIT), the Center for Research in Econometric Analysis of Time Series (CREATES),  the Dale T. Mortensen Center, Aarhus University, and the Danish Council for Independent Research (Grant 0166-00020B) for research support.         }
		\vspace*{1cm}
		
{\normalsize
        \hfill            
{Nicolaj Søndergaard Mühlbach}\symbolfootnote[1]{
             		Department of Economics, MIT, and CREATES.
                Email: \href{mailto:muhlbach@mit.edu}{muhlbach@mit.edu}.}
		\hfill
		}             
 \vspace{1cm}

        {
            \normalsize
            This version: \today
            \par
        }\vspace{-1cm} 

        \thispagestyle{empty}
        \begin{singlespace}
            \begin{abstract}
				\noindent
We propose \OccToVec, a principal approach to representing occupations, which can be used in matching, predictive and causal modeling, and other economic areas. In particular, we use it to score occupations on any definable characteristic of interest, say the degree of \textquote{greenness}. Using more than 17,000 occupation-specific text descriptors, we transform each occupation into a high-dimensional vector using natural language processing.  Similar, we assign a vector to the target characteristic and estimate the occupational degree of this characteristic as the cosine similarity between the vectors. The main advantages of this approach are its universal applicability and verifiability contrary to existing ad-hoc approaches. We extensively validate our approach on several exercises and then use it to estimate the occupational degree of charisma and emotional intelligence (EQ). We find that occupations that score high on these tend to have higher educational requirements.  Turning to wages, highly charismatic occupations are either found in the lower or upper tail in the wage distribution. This is not found for EQ, where higher levels of EQ are generally correlated with higher wages. 			\end{abstract}
        \end{singlespace}
    
	\end{center}
\vspace{0.5cm}
    
\end{titlepage}

\clearpage
\setcounter{page}{1}
\pagenumbering{arabic}

\section{Introduction}\label{sec:Introduction}
Designing optimal education and labor markets,  policy-makers face an increasingly important yet difficult challenge of assessing critical aspects of occupations.  For instance,  \cite{Autor2003,Autor2006,Acemoglu2011,Autor2013skill,Autor2013} find the occupational degree of abstract, routine, and manual job tasks to be important in explaining changes in the labor market; \cite{Goldin2014} examines the role of various occupational characteristics in explaining the gender wage gap; \cite{Deming2017} consider the degree of social skills in occupations; \cite{Frey2017} examine the susceptibility of occupations to computerization; \cite{Brynjolfsson2018a} estimate an occupational suitability index for machine learning (ML); \cite{Bowen2018} examine the \textquote{greenness} of occupations; and \cite{Dingel2020} and \cite{Mongey2021} construct measures of the occupational feasibility of working at home and exposure to social distancing, respectively,  to study the economic impact of social distancing in the perspective of COVID-19.

\VerticalSpace
Common to these seminal studies is the need to measure fundamental characteristics of occupations.  But all existing studies use case-by-case approaches that are deeply difficult to generalize and validate. Most on them rely on detailed data from the Occupational Information Network (O*NET) that provides measures of hundreds of occupational features. Although the richness of the data certainly allows for novel contributions, an inherent risk of overfitting lurks in the dark. That is, researchers might be tempted to select exactly those features of the data that makes the analysis fit the narrative.  The real difficulty is, however, that it may be borderline impossible to validate and verify truly novel measures of occupational characteristics. Another challenge is that every new measure requires a new approach, method, study, etc., which is surely laborious.

\VerticalSpace
We propose \OccToVec, a principal approach to representing occupations as high-dimensional vectors, which can then be used in matching, comparative studies, predictive and causal modeling, and other economic areas of interest. Specifically, we demonstrate how the high-dimensional occupation vectors can be used to score occupations on any definable target characteristic, for instance, the occupational degree of \textquote{greenness}.  At its core, our approach essentially transforms every occupation into a high-dimensional vector for which reason we call it \OccToVec. The only input needed is an objective and reliable textual definition of the target characteristic.  For instance,  the U.S. Bureau of Labor Statistics defines green jobs as Definition \ref{Def:greenness}:
\begin{definition}[Green jobs]\label{Def:greenness}
\textquote{(A) Jobs in businesses that produce goods or provide services that benefit the environment or conserve natural resources. (B) Jobs in which workers’ duties involve making their establishment’s production processes more environmentally friendly or use fewer natural resources.} \citep{bls_greenness}
\end{definition}
Using O*NET, we rely on 244 occupational attributes, 873 occupation descriptions, and 16,804 occupation-specific tasks to learn a vector representation of each occupation, leveraging natural language processing (NLP).\footnote{In principal, any database containing detailed information on the universe of occupations could be used. To our knowledge, O*NET provides the most comprehensive information on occupations. In addition, O*NET provides both textual descriptions and numerical scores, which is necessary for our validation strategy. This will be thoroughly explained in Section \ref{sec:Validation}. An alternative source of occupation data would be the European Skills, Competences, Qualifications and Occupations (ESCO) data, which can be found \href{https://esco.ec.europa.eu/en}{here}. We are not familiar with any papers using the ESCO data, and thus we leave this for further research.} Once all occupations are embedded as high-dimensional vectors, we consider a given target characteristic of interest and assign to it a vector in the same vector space as the occupations. This allows us to estimate the occupational degree of the target characteristic as the (standardized) cosine similarity between its vector and any occupation vector. 

\VerticalSpace
The advantage of this framework is that it is fully data-driven and universal, only requiring the user to provide a reliable and objective definition of the target characteristic. Hence, this is easily extendable to genuinely novel attributes of occupations. In principle, one could extend the framework to default to using an encyclopedia, e.g., Wikipedia,  and then the approach would be truly automated.  Another advantage is how easy it is to validate the framework in contrast to existing approaches that do not consider any ground truth. We validate \OccToVecS in two ways. First, we visually inspect the occupation vectors by compressing them into two-dimensional vectors using principal component analysis (PCA) and $t$-distributed stochastic neighbor embedding ($t$-SNE) \citep{vandermaaten08a}. We plot all occupations on the two dimensions and show that occupations cluster according to major occupational groups and educational requirements.  This indicates that even after compressing the information from the high-dimensional vectors into two dimensions, the low-dimensional occupation vectors still precisely capture differences between occupations. Second, we estimate the occupational degree of all of the 244 occupational attributes from O*NET and compare our estimates to the original O*NET scores. We find strong evidence that our estimates coincide with the original O*NET scores both between and within occupations and take this altogether as evidence that \OccToVecS produces high-quality occupation vectors that capture essential features of the occupations. These vectors can then be used to match occupations more precisely,  act as control covariates in regressions, etc.  Specifically, we demonstrate how the vectors can be used to score occupations on novel characteristics.

\VerticalSpace
Once the quality of the occupation vectors has been scrutinized and confirmed, we showcase the framework using several applications, which we divide into estimating well-studied versus novel characteristics of occupations. 

\VerticalSpace
First, we revisit the popular task measures (abstract, manual, and routine tasks, respectively) by \cite{Autor2003,Autor2006,Acemoglu2011,Autor2013skill,Autor2013} and the innovative measures of artificial intelligence (AI) by \cite{Felten2018},  \cite{Brynjolfsson2018a},  and \cite{Webb2019}.  This illustrates how one could have used our framework had the other case-specific approaches not been invented.  Note carefully that this is not another validation exercise, except under the assumption that the existing measures represent the ground truth. We feel more confident making this assumption for the original O*NET scores.

\VerticalSpace
We document that our estimates of each task measure coincide well with the original measures, and we find that performing a high amount of routine tasks is on average associated with lower educational attainment and lower wages.  We find the same education and wage patterns for occupations that are characterized by performing manual tasks, whereas the opposite holds for abstract tasks.

\VerticalSpace
We find that our estimates of AI exposure generally behave as a mix of the measures of \cite{Felten2018} and \cite{Webb2019} with no clear relationship to that of \cite{Brynjolfsson2018a}. The close similarity of \cite{Felten2018} and \cite{Webb2019} in contrast to \cite{Brynjolfsson2018a} was first noted by \cite{Acemoglu2022ai}. Occupations that have a high degree of AI have similar profiles as those that score high on abstract tasks. That is, these occupations generally have higher educational requirements and enjoy higher wages.

\VerticalSpace
Second, we investigate the occupational degree of two novel characteristics, namely charisma and emotional intelligence (EQ). We use definitions from psychological outlets, e.g., the American Psychological Association. Both characteristics are intrinsically interesting and have been found to play a fundamental role in leadership skills and social affability.  At the outset, charisma and EQ seem conceptually closely related, and we do find similarities empirically.  The occupations that score high on both charisma and EQ are often found within community and social service,  educational instruction, and arts,  design, entertainment, sports, and media occupations. as well as educational instruction.  For charisma,  another frequent occupational group is sales. The occupations with high degrees of charisma and EQ also share similar wage profiles. Particularly, our estimates suggest occupations in the upper part of the wage distribution score high on both charisma and EQ. One difference appears in the lower part of the wage distribution, where occupations tend to score high on charisma but not on EQ. Another difference that we detect at the aggregate level enters between occupations that require master's degree versus doctoral degrees.  Essentially, these occupations score similarly on charisma, whereas there is a dip in the level of EQ for occupations that require doctorates.

\VerticalSpace
The rest of the paper is organized as follows. Section \ref{sec:Data} presents the data used in \OccToVec, whereas Section \ref{sec:Methodology} introduces the framework and our NLP method of choice. Section \ref{sec:Validation} validates our framework and Section \ref{sec:Application} considers our four applications on task measures, AI, charisma, and EQ, respectively. Section \ref{sec:Conclusion} concludes.  Appendices to data, method, and applications are found in Appendix \ref{App:data}, \ref{App:method}, and \ref{App:tasks}--\ref{App:eq}, respectively. Throughout the paper we will be using the notation $\left|A\right|$ for the cardinality of a generic set $A$ and $\left[a,b\right]=\left\{ z\in\mathbb{Z}\vert a\leq z\leq b\right\}$ denotes the set of integers between $a$ and $b$ with $\left[a\right]$ simply meaning $\left\{ 1,\ldots,a\right\}$.

\section{Data}\label{sec:Data}
In this section, we describe the data used to construct the occupation vectors. Starting with the data has two advantages. First, in our experience, it is easier to understand the framework with a specific source of data in mind, especially for readers less versed in NLP.  Second,  although any detailed occupational database could be used, O*NET provides the most comprehensive information to our knowledge. Specifically, the validation strategy requires the source of occupational data to contain both textual descriptions and numerical scores for descriptors of the entire universe of occupations, which we have not come across elsewhere.\footnote{Specifically, the O*NET\textsuperscript \textregistered \SpaceBar Content Model, which can be found \href{https://www.onetcenter.org/content.html}{here}. We use O*NET version 26.3, which can be found \href{https://www.onetcenter.org/dictionary/26.3/excel/}{here}.}

\VerticalSpace
O*NET keeps track of hundreds of descriptors for each occupation and we use the 873 unique occupations available.  Each occupation is described textually by a general \textit{description} (e.g.,  writers generally originate and prepare written material, such as scripts, stories, advertisements, and other material) and by detailed \textit{tasks} (e.g., bakers place dough in pans, molds, or on sheets), as well as measured numerically on several \textit{attributes} (e.g.,  how important is critical thinking for surgeons). We denote the three sources of occupational information (i.e.,  descriptions, tasks, and attributes) collectively as occupational \textit{descriptors}.  We highlight this distinction because the majority of the existing approaches is only suited for using the numerical scores of the attributes, whereas our method incorporates purely textual data as the tasks and descriptions as well.\footnote{One important exception is \cite{Webb2019}, who uses the tasks but not the attributes.  To the best of our knowledge,  \cite{Webb2019} cannot be generalized to using the attributes as well.} Note that attributes are also defined textually and that tasks are also associated with a weight (i.e., a numerical score). Thus,  all the occupational descriptors are both expressed as text and associated with an occupation-specific numerical score.  For instance, the definitions of the attributes \textit{oral comprehension} and \textit{depth perception} follow from Table \ref{Tbl:onet_attribute_example}.  These textual definitions naturally apply to all occupations, but each occupation is also assigned a specific score between zero and one for all attributes. For instance, being an astronomer is associated with a score of 0.74 on oral comprehension and of 0.25 on depth perception.

\begin{table}[!t]
\caption{Two examples of O*NET attributes}
	\begin{adjustbox}{max totalsize = {\textwidth}{0.9\textheight}, center}
\begin{tabular}{p{0.15\linewidth}p{0.25\linewidth}p{0.7\linewidth}}
\toprule 
Element ID & Element name & Definition\tabularnewline
\midrule
1.A.1.a.1 & Oral Comprehension & The ability to listen to and understand information and ideas presented through spoken words and sentences.\tabularnewline
1.A.4.a.6 & Depth Perception & The ability to judge which of several objects is closer or farther away from you, or to judge the distance between you and an object.\tabularnewline
\bottomrule
\end{tabular}
	\end{adjustbox}\VerticalSpaceFloat
\label{Tbl:onet_attribute_example}
{\footnotesize
\textit{Notes:} This table shows the definition of two occupational attributes from the O*NET\textsuperscript \textregistered \SpaceBar Content Model. The examples include the ability \textit{oral comprehension} (\href{https://www.onetonline.org/find/descriptor/result/1.A.1.a.1}{link}), and the ability \textit{depth perception} (\href{https://www.onetonline.org/find/descriptor/result/1.A.4.a.6}{link}).\par}
\end{table}

Similar, Table \ref{Tbl:onet_occupation_example} exemplifies an occupation description as well as a few tasks for astronomers.  Additionally, the descriptors are associated with a weight.

\begin{table}[!t]
\caption{Example of occupation description and tasks}
	\begin{adjustbox}{max totalsize = {\textwidth}{0.9\textheight}, center}
\begin{tabular}{p{0.15\linewidth}p{0.8\linewidth}p{0.15\linewidth}}
\toprule 
Category & Definition & Weight\tabularnewline
\midrule
Description & Observe, research, and interpret astronomical phenomena to \newline increase basic knowledge or apply such information to practical problems. & 1\tabularnewline
Task & Calculate orbits and determine sizes, shapes, brightness, and motions of different celestial bodies.  & 0.053\tabularnewline
Task & Measure radio, infrared, gamma, and x-ray emissions from \newline extraterrestrial sources. & 0.061\tabularnewline
\bottomrule
\end{tabular}
	\end{adjustbox}\VerticalSpaceFloat
\label{Tbl:onet_occupation_example}
{\footnotesize
\textit{Notes:} This table shows the description and two tasks of astronomers from the O*NET\textsuperscript \textregistered \SpaceBar Content Model.  The summary report for astronomers can be found \href{https://www.onetonline.org/link/summary/19-2011.00}{here}.\par}
\end{table}

\begin{figure}[!t]
\begin{adjustbox}{max totalsize = {\textwidth}{\SingleFigMaxHeight\textheight}, center}
\includegraphics[width = \textwidth]{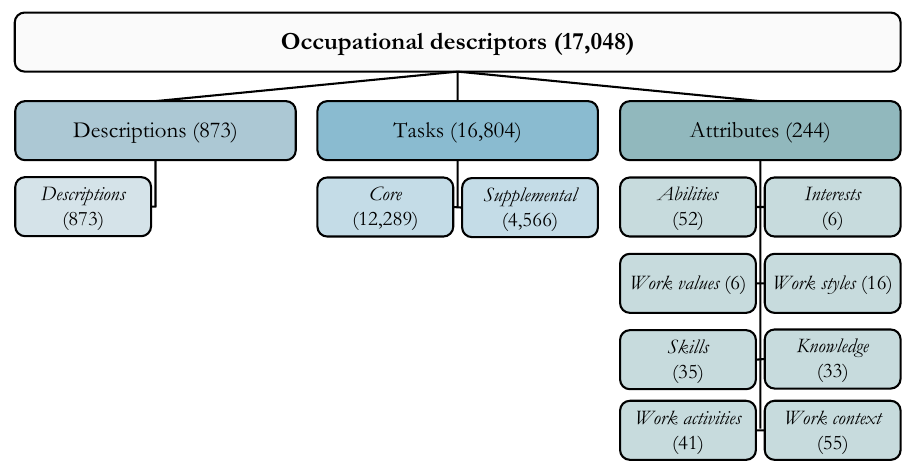}
\end{adjustbox}	
\caption{Overview of the occupational descriptors and subcategories}
\label{Fig:onet_data}
\VerticalSpaceFloat
{\footnotesize
\textit{Notes:} This figure shows the occupational descriptors provided by O*NET that can be expressed meaningfully by both text and numeric scores. The 17,048 unique occupational descriptors consist of 873 occupation descriptions,  16,804 detailed tasks, and 244 attributes. Note that a small number of tasks may be core to some occupations while supplemental to others, meaning that the sum of core and supplemental tasks exceed the total number of unique tasks.  We divide the descriptors into ten categories of which one category belongs to descriptions,  one category belongs to tasks (core and supplemental tasks are grouped as one category), and eight categories belong to attributes as it appears on the figure.\par}
\end{figure}

\VerticalSpace
The occupational descriptors represent ten categories; namely, \textit{description}, \textit{tasks}, \textit{abilities}, \textit{interests}, \textit{work values}, \textit{work styles}, \textit{skills}, \textit{knowledge}, \textit{work activities},  and \textit{work context}.\footnote{Technically, the ten categories further represent four broader types of information about work; worker characteristics, worker requirements, occupational requirements, and occupation-specific information. Specifically, worker characteristics include abilities, interests, work values,  and work styles; worker requirements include skills and knowledge; occupational requirements include work activities and work context; and occupation-specific information include description and tasks. In this paper, we focus on the ten categories and we do not distinguish further between the four broader types of occupational information.} The total number of occupational descriptors amounts to 17,048, which includes 873 occupation descriptions (one for each occupation), 16,804 unique tasks (some tasks are occupation-specific, while others are performed by a few occupations, leading to on average 20 tasks per occupation), and 244 attributes (common across occupations but with occupation-specific weights). Figure \ref{Fig:onet_data} shows the structure of the data.

We are exhaustive in the selection of occupational descriptors in the sense that we include all those for which meaningful textual descriptions are included.\footnote{For instance, the ability \textit{stamina} is defined by \textit{the ability to exert yourself physically over long periods of time without getting winded or out of breath},  whereas the technological skill \textit{electronic mail software} (e.g., Microsoft Outlook) is not further defined.} The occupation-specific weight that is associated with each descriptor is constructed based on at least one scale, e.g., importance, level, etc., and the realized value of the scale differs by occupations.  The definition of each scale along with the set of categories that each scale applies to are presented in Table \ref{Tbl:onet_scales} in Appendix \ref{App:data}.  Each scale has a minimum and maximum value and to enable comparison, all scales have been standardized to ranging from 0 to 1 following O*NET guidelines.  In the case of multiple scales for a specific descriptor, we take a uniform average of the standardized scales. 
\section{Methodology}\label{sec:Methodology}
In this section, we present our principal approach to measuring any definable target characteristic of an occupation by quantifying the similarity between two vectors that represent the occupation and the target characteristic, respectively. The occupation vectors are text \textit{embeddings} (more intuition below) generated by our preferred NLP method that heavily rely on the plethora of occupation-specific information provided by the O*NET, and thus, our approach essentially transforms every occupation into a high-dimensional vector that captures the tasks and attributes of the occupation. Similar,  the vector of the target characteristic is a text embedding of its definition that can in principle be provided by any objective and reliable source of information, e.g.,  international standards or dictionaries.  The characteristic of interest, say the \textquote{greenness} of occupations,  is thereby also transformed into a high-dimensional vector, which allow us to compute a measure of similarity of the two vectors; one vector representing a particular occupation and one vector representing the characteristic.

\subsection{Intuition behind text embeddings}
Before deep-diving into details, we provide some intuition of the text embeddings that are the central building blocks of \OccToVec. Overall speaking,  a text embedding is a real-valued vector that represents a mathematical analogue of the piece of text in question such as words, sentences,  paragraphs, etc.  The vector encodes the meaning of the text such that the texts that are closer in the vector space are expected to be similar in meaning. In a highly-simplified one-dimensional world, this would mean that the occupation ``actuary'' could have a mathematical representation of, say, the number 2 and ``mathematician'' the number 3. This means that linguistically  ``actuary'' and ``mathematician'' are similar in the sense that their tasks and attributes overlap to a high degree. In contrast, imagine ``actors'' and ``dancers'' would be represented by, say,  11 and 14, respectively, meaning that actors and dancers are more similar to each other than to either actuaries or mathematicians.\footnote{In modern applications of text embeddings, the dimensions are several hundreds and we use $p=1024$.} Thus,  the process of mapping pieces of texts to real numbers leads to the creation of embeddings that capture the meaning of the text. The benefit of using text embeddings is that it allows texts with similar meanings to have similar vector representations (as measured by high cosine similarity).

\VerticalSpace
Embedding more than 17,000 textual descriptors, plenty of NLP methods exist. Common to all is the assumption of the Distributional Hypothesis due to \cite{Harris1954}, claiming that words that occur in the same contexts tend to have similar meanings and also popularized by \cite{Firth1957} as \textquote{\textit{a word is characterized by the company it keeps}}. Some of the natural choices to consider are Word2Vec \citep{Mikolov2013EfficientEO}, its document version Doc2Vec/Paragraph2Vec \citep{Quoc2014},  GloVe \citep{Pennington2014},  Fasttext \citep{Bojanowski2016,Joulin2016a,Joulin2016b}, or BERT \citep{Devlin2018}. In this paper, we use an optimized version of BERT called Sentence-RoBERTa \citep{reimers2019,Liu2019}, and thus our foundation is BERT. Our framework is not limited to specific embedding techniques, and our findings are robust to the choice of NLP methods. We will not review all details of BERT/RoBERTa as this is out of scope. Instead, we provide an intuitive explanation of BERT/RoBERTa and refer readers to Appendix \ref{App:method} for more details or \cite{Rogers2021} for an excellent review.

\VerticalSpace
Let $T$ be a given piece of text (e.g., a sentence or a paragraph) from corpus $\mathcal{T}$. Each $T$ is essentially a sequence of tokens (e.g., words or subwords).  Further,  let $D\in\mathbb{R}^{p}$ be an embedding of $T$, i.e., a $p$-dimensional vector that mathematically represents the meaning of $T$.  At least conceptually, the assumption underlying all approaches to text embeddings is that there exists a map $f:T\rightarrow D$ approximating the relationship between $T$ and $D$. Any text embedding model may be viewed as a nonparametric estimate $\hat{f}$ of the map $f$. The challenge is that the true embeddings are latent, and thus not observed. The various methods consider different approaches to circumventing this, but the most common routine to learn $f$ is to mask a random sample of tokens and then train the model to produce embeddings by minimizing the cross-entropy loss from predicting the masked tokens given the estimated embeddings, which is essentially a cloze procedure \citep{Taylor_1953}. The training process encodes a lot of semantics and syntactic information about the language by training the model on massive amounts of unlabeled textual data drawn from the web. The objective is to detect generic linguistic patterns in the text.

\subsection{General framework}
Index occupations by $i$ for $i\in\left[n\right]$ and assume access to data $\left\{ \left(W_{j},T_{j}\right):j\in\left[d\right]\right\}$, where $T_{j} \in\mathcal{T}$ is the textual definition of the $j$th occupational descriptor,  $\mathcal{T}$ represents the universe of unique descriptor definitions, and $W_{j}\in\left[0,1\right]^{n}$ is an $n$-dimensional vector of occupation weights on the $j$th descriptor such that $w_{i,j}$ represents the importance of descriptor $j$ for occupation $i$.  Let $\mathbf{W}=\left(W_{1}\cdots W_{d}\right)\in\left[0,1\right]^{n\times d}$ denote the matrix of concatenated weights.\footnote{We normalize $\mathbf{W}$ such that each $\mathbf{W}_{i,\cdot}$ sum to unity for $i\in\left[n\right]$. We further normalize $\mathbf{W}$ such that each of the ten O*NET categories contribute equally, that is, weights representing the same category sum to one over the number of categories.} In principle, all occupations could be represented by any descriptor but in practice many weights are zero because the descriptors contain tasks only performed by a subset of occupations.

\VerticalSpace
The goal of the NLP algorithm of choice is to turn each descriptor definition $T_{j}$ into a descriptor embedding $D_{j}\in\mathbb{R}^{p}$, where $p$ represents the dimensionality of the embeddings, leading to an embedding matrix $\mathbf{D}=\left(D_{1}\cdots D_{d}\right)\in\mathbb{R}^{d\times p}$. The occupation embeddings are then constructed as a weighted average of the descriptor embeddings via \eqref{Eq:embedding_mechanism},  i.e.
\begin{align}\label{Eq:embedding_mechanism}
	\mathbf{X}=\mathbf{W}\mathbf{D},
\end{align}
where $\mathbf{X}=\left(X_{1} \cdots X{}_{n}\right)\in\mathbb{R}^{n\times p}$, and $X_i \in\mathbb{R}^{p}$ represents occupation $i$ as a $p$-dimensional vector. Note that $p$ is typically several hundreds and we use $p=1024$.  Using this representation of occupations, we think of a given occupation as a weighted average of its descriptors (attributes, tasks, and title), where the weights are governed by the combined importance, relevance, and frequency of the descriptor. 

\subsubsection{Fine-tuning occupation vectors}
Tailoring \OccToVecS to occupational data, we develop a novel fine-tuning step.\footnote{Since we only have access to 873 occupations via O*NET, we cannot fine-tune the embeddings in the standard way of formulating a deep neural network to classify the occupations.} Imagine access to a new, unseen target occupational characteristic, $T_0$, and an $n$-dimensional vector $Y_0$ of occupational scores on $T_0$.  Similar to the construction of $\mathbf{D}$, we use our NLP algorithm to embed $T_0$ into a $p$-dimensional vector $D_0$. Our goal is then to fine-tune the occupation embeddings, $\mathbf{X}$, such that the cosine similarity between the occupation embeddings and the target characteristic embedding predicts the score, $Y_0$, as closely as possible.  The cosine similarity between a particular occupation embedding $X_i$ and the target characteristic embedding $D_0$ is given by
\begin{align}\label{Eq:cosine_similarity}
S_{\cos}\left(X_{i},D_{0}\right)=\frac{X_{i}\cdot D_{0}}{\left\Vert X_{i}\right\Vert _{2}\left\Vert D_{0}\right\Vert _{2}}=\frac{\sum_{k=1}^{p}X_{i,k}D_{0,k}}{\sqrt{\sum_{k=1}^{p}X_{i,k}}\sqrt{\sum_{k=1}^{p}D_{0,k}}},
\end{align}
which simplifies to the dot product, $S_{\cos}\left(X_{i},D_{0}\right)=X_{i}\cdot D_{0}$, because both embeddings are normalized to unit length. Our approach to fine-tuning the occupation embeddings is to estimate a $\left(p\times p\right)$-dimensional rotation matrix $\mathbf{R}$ that solves
\begin{align}\label{Eq:R_minimization_unseen}
	\mathbf{R} &= \arg\min\left(\left\Vert Y_0-\mathbf{X} \mathbf{R} D_0\right\Vert _{2}^{2}\right)
\end{align}
Estimating $\mathbf{R}$ via Eq. \eqref{Eq:R_minimization_unseen}, it is most appropriate\footnote{Technically, we avoid severely sparse matrices by focusing on common descriptors.} to consider descriptors that are shared across all occupations, namely the attributes in the case of O*NET. For this reason, let $\mathcal{D}$ denote the entire set of descriptors and let $\mathcal{A}$ denote the (smaller) set of attributes that are shared among occupations, such that $\mathcal{D}\setminus\mathcal{A}$ represents tasks and job titles. Then, let $\mathbf{Y}=\mathbf{W}_{\cdot,\left\{ j:j\in\mathcal{A}\right\} }\in\left[0,1\right]^{n\times a}$ be the subset of columns in $\mathbf{W}$ that represents the attributes among all descriptors, where $a=\left|\mathcal{A}\right|$.\footnote{Recall that descriptors include attributes (shared among occupations), tasks (unique to subset of occupations), and titles (unique per occupation).} We interpret each entry in $\mathbf{Y}$, namely $y_{i,j}$ for $i\in\left[n\right]$ and $j\in \mathcal{A}$,  as representing the score of occupation $i$ on attribute $j$. Similarly, let $\mathbf{A}=\mathbf{D}_{\left\{ i:i\in\mathcal{A}\right\} ,\cdot}\in\mathbb{R}^{a\times p}$ denote the subset of rows in $\mathbf{D}$ that represent the attributes, such that $\mathbf{A}_{j,\cdot}$ represents the $p$-dimensional vector embedding of attribute $j$.  The empirical matrix analog to \eqref{Eq:R_minimization_unseen} then reads
\begin{align}\label{Eq:R_unconstrained}
	\mathbf{R}_\text{UR} &= \arg\min\left(\left\Vert \mathbf{Y}-\mathbf{X} \mathbf{R} \mathbf{A}'\right\Vert _{F}^{2}\right),
\end{align}
where subscript UR abbreviates unrestricted,  and $\left\Vert \cdot \right\Vert _{F}$ denotes the Frobenius norm.  The optimization problem in Eq.  \eqref{Eq:R_unconstrained} without any constraints has an analytic solution given by 
\begin{align}\label{Eq:R_optim_no_constraints}
\mathbf{R}_\text{UR}=\left(\mathbf{X}'\mathbf{X}\right)^{-1}\mathbf{X}'\mathbf{Y}\mathbf{A}\left(\mathbf{A}'\mathbf{A}\right)^{-1},
\end{align}
provided that $\mathbf{X}$ and $\mathbf{A}$ have full column rank such that the inverses exist. 

\VerticalSpace
Empirically, we find, however, that using Eq. \eqref{Eq:R_unconstrained} rather than imposing additional constraints leads to sub-optimal performance on unseen data.\footnote{Essentially, setting $\mathbf{R}=\mathbf{I}_{p}$ performs comparably to $\mathbf{R}_\text{UR}$ on data not used in the estimation of $\mathbf{R}_\text{UR}$} This is most likely due to overfitting as the unconstrained optimization problem is highly overparametrized. Thus, we experiment with three additional constraints, namely $\mathbf{R}_{k,k'\in\left[p\right]:k,k'\in k\neq k'}=0$ (ensuring that $\mathbf{R}$ is diagonal), $\mathbf{R}_{k,k'\in\left[p\right]}\geq0$ (ensuring that all entries in $\mathbf{R}$ are non-negative), and $\sum_{k}\sum_{k'}\mathbf{R}_{k,k'}\leq c$, where $c$ is some constant (ensuring that the sum of entries in $\mathbf{R}$ is constrained to avoid too much flexibility). We always enforce $\mathbf{R}$ to be diagonal and refer to the rotation matrices stemming from each constraint as $\mathbf{R}_\text{DIAG}$, $\mathbf{R}_\text{DIAG+NONNEG}$, and $\mathbf{R}_\text{DIAG+SUM}$, respectively. To no-tuning benchmark is $\mathbf{R}=\mathbf{I}_{p}$, where $\mathbf{I}_{p}$ is the $p$-dimensional identity matrix, which we denote $\mathbf{R}_\text{IDENTITY}$.

\VerticalSpace
To avoid overfitting, we never re-use the same data to estimate the occupation vectors in $\mathbf{X}$ and the rotation matrix $\mathbf{R}$, or to evaluate the performance. Hence,  we randomly split the set of attributes multiple times into three equal-sized sets without replacement such that $\mathcal{A}=\mathcal{A}_{1}\cup\mathcal{A}_{2}\cup\mathcal{A}_{3}$ and $\mathcal{A}_{h}\cap\mathcal{A}_{g}=\emptyset\;\forall h\neq g$. Then, define $\mathbf{Y}^{g}=\mathbf{Y}_{\cdot,\left\{ j:j\in\mathcal{A}_{g}\right\} }$ for $g=1,2,3$, such that the concatenation $\tilde{\mathbf{Y}}=\left(\mathbf{Y}^{1},\mathbf{Y}^{2},\mathbf{Y}^{3}\right)$ is simply a reshuffled version of $\mathbf{Y}$. Similarly,  define $\mathbf{A}^{g}=\mathbf{A}_{\cdot,\left\{ j:j\in\mathcal{A}_{g}\right\} }$ for $g=1,2,3$. Last, define $\mathbf{W}^{g}=\mathbf{W}_{\cdot,\left\{ j:j\notin\cup_{h\neq g}\mathcal{A}_{h}\right\}}$ and $\mathbf{D}^{g}=\mathbf{D}_{\left\{ i:i\notin\cup_{h\neq g}\mathcal{A}_{h}\right\} ,\cdot}$ for $g=1,2,3$. Having the definitions in place, we compare the various strategies for obtaining an estimate of $\mathbf{R}$ by the following procedure:
\begin{enumerate}\label{List:estimation_procedure}
	\item Compute occupational embeddings $\mathbf{X}^{1}=\mathbf{W}^{1}\mathbf{D}^{1}$
	\item Estimate rotation matrix $\mathbf{R}^{2}=\arg\min_{\mathbf{R}}\left(\left\Vert \mathbf{Y}^{2}-\mathbf{X}^{1}\mathbf{R}\mathbf{A}^{2'}\right\Vert _{F}^{2}\right)$ potentially subject to constraints
	\item Evaluate performance by $\mathbf{Q}^{3}=\left\Vert \mathbf{Y}^{3}-\mathbf{X}^{1}\mathbf{R}^{2}\mathbf{A}^{3'}\right\Vert _{F}^{2}$
\end{enumerate}
We re-do the above procedure five times (with a random split of attributes each time) and report the performance relative to the benchmark in Table \ref{Tbl:assessment_R_estimation}.

\begin{table}[!t]
\caption{Performance overview of tuning strategies}
	\begin{adjustbox}{max totalsize = {\textwidth}{0.9\textheight}, center}
		\begin{tabular}{llrrrrrr}
\toprule
              &          & \multicolumn{3}{l}{Frobenius norm} & \multicolumn{3}{l}{Nuclear norm} \\
              &          &        average &   max &   min &      average &   max &   min \\
Sample & Strategy &                &       &       &              &       &       \\
\midrule
Full sample & IDENTITY &          1.000 & 1.000 & 1.000 &        1.000 & 1.000 & 1.000 \\
              & UR &          0.598 & 0.626 & 0.585 &        0.483 & 0.497 & 0.475 \\
              & DIAG &          0.739 & 0.752 & 0.729 &        0.961 & 0.963 & 0.958 \\
              & DIAG+NONNEG &          0.742 & 0.745 & 0.740 &        0.946 & 0.945 & 0.946 \\
              & DIAG+SUM &          0.754 & 0.762 & 0.748 &        0.978 & 0.980 & 0.977 \\
In-Sample & IDENTITY &          1.000 & 1.000 & 1.000 &        1.000 & 1.000 & 1.000 \\
              & UR &          0.000 & 0.000 & 0.000 &        0.000 & 0.000 & 0.000 \\
              & DIAG &          0.580 & 0.595 & 0.582 &        0.871 & 0.873 & 0.878 \\
              & DIAG+NONNEG &          0.680 & 0.689 & 0.678 &        0.912 & 0.914 & 0.916 \\
              & DIAG+SUM &          0.598 & 0.613 & 0.603 &        0.893 & 0.894 & 0.896 \\
Out-of-Sample & IDENTITY &          1.000 & 1.000 & 1.000 &        1.000 & 1.000 & 1.000 \\
              & UR &          1.006 & 1.023 & 1.019 &        1.149 & 1.134 & 1.182 \\
              & DIAG &          0.964 & 0.980 & 0.951 &        1.012 & 1.002 & 1.008 \\
              & DIAG+NONNEG &          0.842 & 0.857 & 0.833 &        0.958 & 0.949 & 0.955 \\
              & DIAG+SUM &          0.977 & 0.978 & 0.959 &        1.027 & 1.019 & 1.026 \\
\bottomrule
\end{tabular}

	\end{adjustbox}\VerticalSpaceFloat
\label{Tbl:assessment_R_estimation}
{\footnotesize
\textit{Notes:} This table shows the relative performance of various tuning strategies relative to the no-tuning benchmark, where the rotation matrix equals the identity matrix. We report the results in terms of the average, minimum, and maximum Frobenius and Nuclear norm across the five random splits, when the tuning strategies are evaluated on the entire sample,  in-sample, and out-of-sample, respectively.\par}
\end{table}

Since our objective is to generalize the performance of our occupation embeddings to unseen target characteristics, the most interesting rows are in the bottom of Table \ref{Tbl:assessment_R_estimation}, representing the out-of-sample performance.  The best performing strategy on both loss functions is $\mathbf{R}_\text{DIAG+NONNEG}$, where the rotation matrix is forced to be diagonal and have non-negative entries. The gain relative to the benchmark is almost 16\% in terms of the Frobenius norm, which we optimize with respect to. Note also from Table \ref{Tbl:assessment_R_estimation} how especially $\mathbf{R}_\text{UR}$ is prone to overfitting since the in-sample performance is essentially perfect but the out-of-sample performance is actually marginally worse than the benchmark.  Thus, we overwrite the occupation vectors and shall henceforth use $\mathbf{X}\coloneqq\mathbf{X}\mathbf{R}_{\text{DIAG+NONNEG}}$. 
\section{Validation}\label{sec:Validation}
In this section, we validate \OccToVecS by performing two exercises. The first exercise is exploratory and visual, where we show that occupation vectors can be well separated by major occupational groups and required educational attainment. The second exercise is more formal and statistical, where we use various metrics to assess the performance of our estimates of the same occupational attributes that O*NET provides. Note that if one is interested in characteristics that appear in the O*NET database (any of the 244 occupational attributes), there is no reason to use our approach.  In this case, one should use the O*NET scores directly. Our method is relevant when one is considering novel characteristics that are not included in O*NET.  Taking the O*NET scores as the ground truth simply allows for an transparent validation of our approach.

\subsection{Visualizing occupations}
\begin{figure}[!t]
\begin{adjustbox}{max totalsize = {\textwidth}{0.4\textheight}, center}
\includegraphics[width = \textwidth]{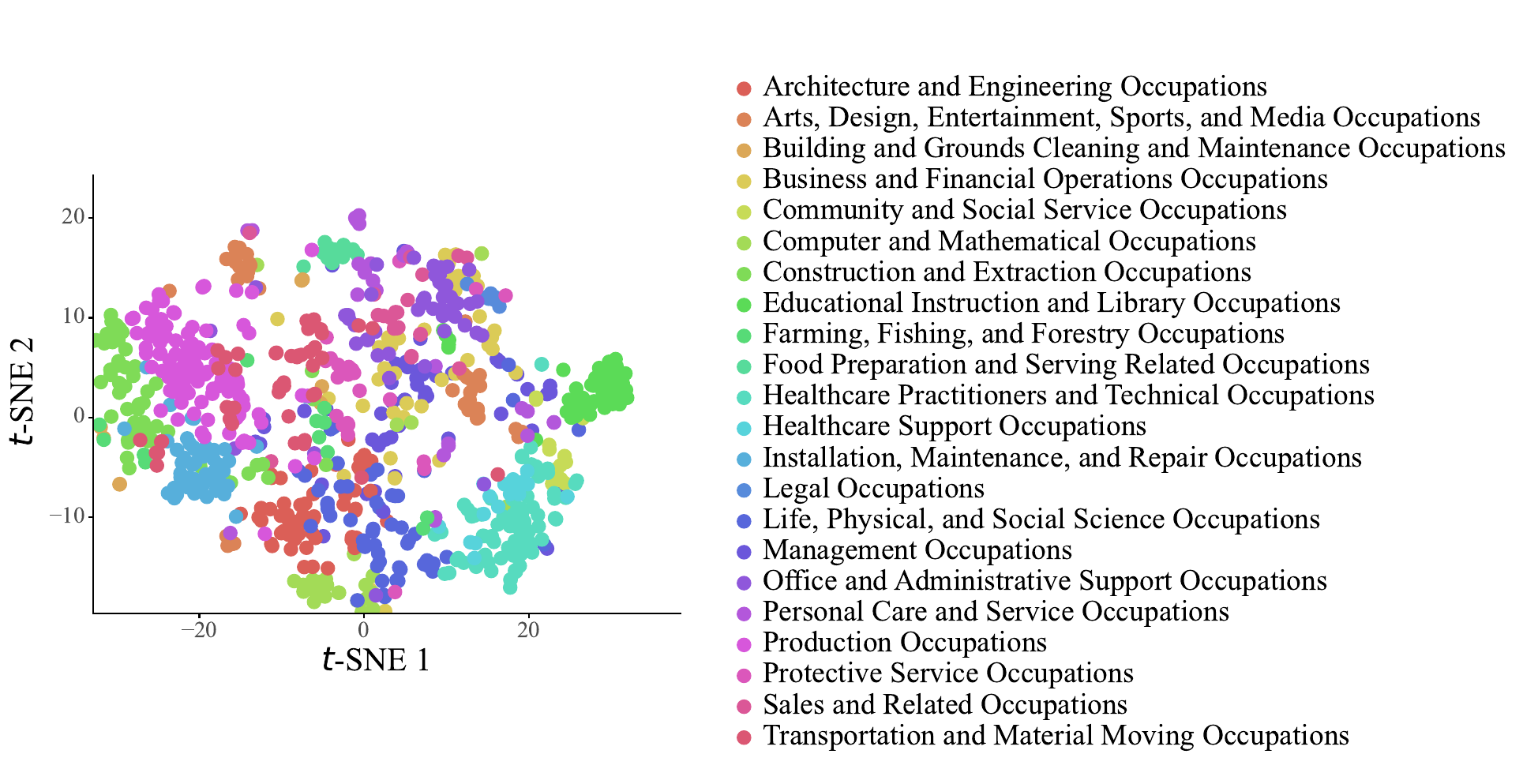}
\end{adjustbox}	
\caption{Occupation embeddings by major occupation groups}
\label{Fig:2d_visualizing_embeddings_by_major_group}
\VerticalSpaceFloat
{\footnotesize
\textit{Notes:} This figure shows the occupational embeddings in a two-dimensional space, where each dot represents a unique occupation. We apply PCA and $t$-SNE to reduce the dimensionality of the embeddings to two dimensions.  The occupations are color-coded by the corresponding major occupational group.\par}
\end{figure}

\begin{figure}[!t]
\begin{adjustbox}{max totalsize = {\textwidth}{\SingleFigMaxHeight\textheight}, center}
\includegraphics[width = \textwidth]{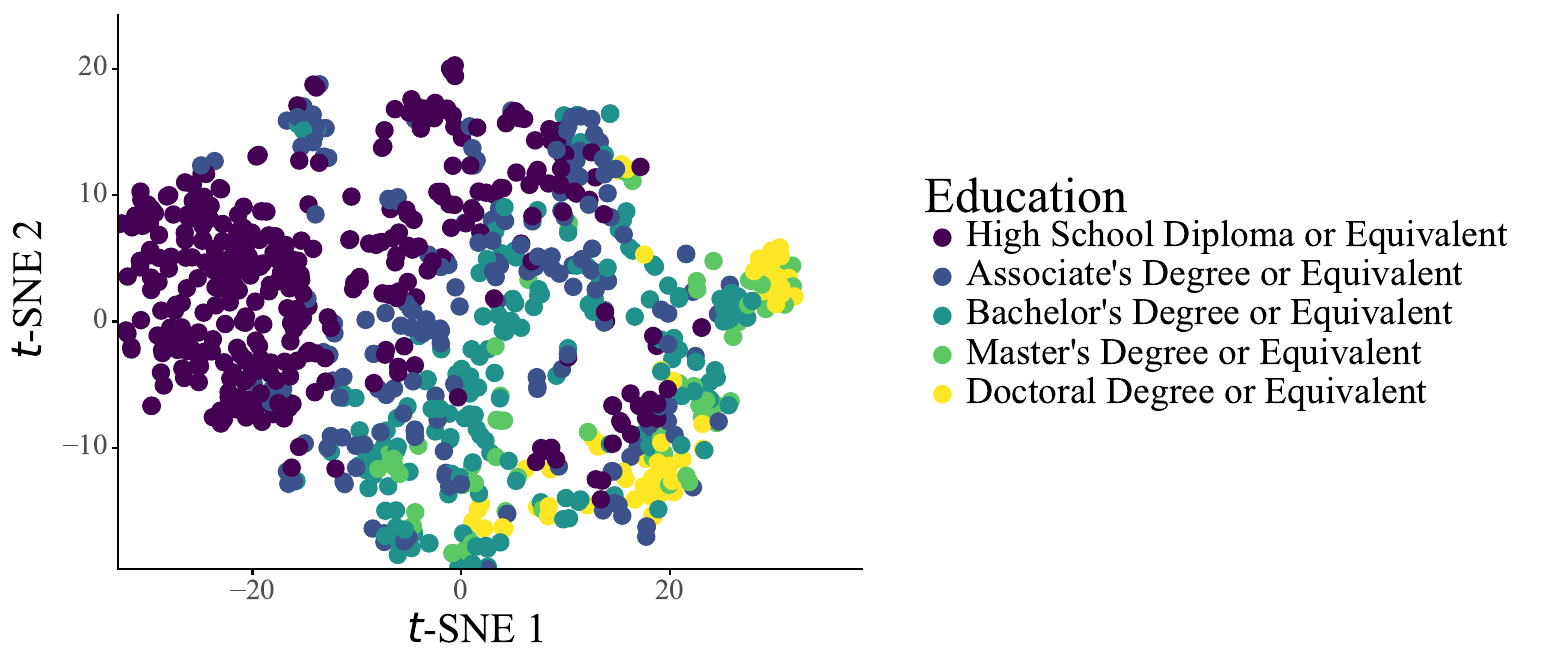}
\end{adjustbox}	
\caption{Occupation embeddings by education}
\label{Fig:2d_visualizing_embeddings_by_education}
\VerticalSpaceFloat
{\footnotesize
\textit{Notes:} This figure shows the occupational embeddings in a two-dimensional space, where each dot represents a unique occupation. We apply PCA and $t$-SNE to reduce the dimensionality of the embeddings to two dimensions.  The occupations are color-coded by the corresponding educational requirement. \par}
\end{figure}

Having outlined how we quantify an occupation by transforming it to a high-dimensional vector, we illustrate the occupations in a two-dimensional space and color them by major occupational groups and educational requirements in Figure \ref{Fig:2d_visualizing_embeddings_by_major_group} and \ref{Fig:2d_visualizing_embeddings_by_education}, respectively. Reducing the 1024 dimensions of the embeddings to only two dimensions, we first use PCA to reduce the number of dimensions to 50 and then we use $t$-SNE \citep{vandermaaten08a} to further reduce the dimensionality to two.\footnote{This is a popular tool to visualize high-dimensional data, which works by converting similarities between data points to joint probabilities and minimizing the Kullback-Leibler divergence between the joint probabilities of the low-dimensional embedding and the high-dimensional data. }

\VerticalSpace
From Figure \ref{Fig:2d_visualizing_embeddings_by_major_group},  it follows that the 873 occupations can robustly be separated into the 22 major occupational groups as the occupations tend to cluster into these groups (of the same color).  Note that the different colors have no interpretation as groups are sorted alphabetically. For instance,  both production occupations in purple,  construction and extraction occupations in green, and installation, maintenance, and repair occupations in blue can be found in the LHS, and despite the differences in colors, these groups represent a cluster related to physical and manual occupations.

\VerticalSpace
Likewise, Figure \ref{Fig:2d_visualizing_embeddings_by_education} indicates that educational requirement is an important separator of occupations. Especially the two extremes, i.e.,  high school diploma and doctoral degree, occupy two different positions in the two-dimensional space. Note that the colors do have an interpretation is this figure, because we can order educations by level. 

\VerticalSpace
The two figures support the idea that occupations are quantifiable and separable and this is despite the fact that we compress all the information in the high-dimensional space to only two dimensions. When comparing the figures, note that they would be identical without the coloring because we only perform the dimension reduction once. 

\subsection{Comparing \OccToVecS estimates to the original O*NET scores}
Recall that O*NET provides 244 occupational attributes for each of the 873 occupations.  Principally, we may estimate the occupational degree of each of these attributes the same way as we would for any target characteristic of interest.  That is, we compare the occupation embeddings to each of the attribute embeddings, thereby quantifying the degree of each attribute as the cosine similarity. This gives us 244 estimates of occupational attributes for each of the 873 occupations, totaling 213,012 estimates we can readily use to validate our framework assuming O*NET to be the the ground truth. This is possible because O*NET provides both numerical scores as well as textual definitions for all attributes.\footnote{Note that we cannot do this for the 873 occupation descriptions nor the 16,804 occupation-specific tasks both because these descriptors not comparable across occupations.}

\paragraph*{Between and within occupations}
First, we assess the ability to correctly estimate an occupational attribute by considering correlations \textit{between} and \textit{within} occupations. The reason we care about a high correlation between occupations is that for a specific attribute, we should be able to rank the occupations accordingly (e.g., how do judges compare to carpenters on critical thinking?). To assess between-occupations validity, we use the 213,012 attribute estimates and compute the Spearman correlation coefficient between the O*NET score and our estimate by attribute, yielding 244 correlation estimates. 

\VerticalSpace
A high correlation within occupations is equally important because for a specific occupation, we should be able to rank multiple attributes (e.g., how does static strength compare to dynamic strength for firefighters?). To assess within-occupations validity, we again use the 213,012 attribute estimates and compute the Spearman correlation coefficient but this time by occupation, yielding 873 correlation estimates.

\VerticalSpace
Obtaining two sets of correlations, we determine if each mean correlation differs significantly from zero using classic $t$-tests.  In additional,  we test if each mean correlation differs significantly from hypothesized values ordered from 1\% to 99\% and report the first postulated correlation for which we fail to reject the null hypothesis of no significant difference at the 5\% level. The results follow from Table \ref{Tbl:correlation_between_measures}, where Table \ref{Tbl:correlation_between_occupations} shows the between-occupations tests and Table \ref{Tbl:correlation_within_occupations} shows the within-occupations tests. 

\VerticalSpace
Both types of validity tests strongly reject the null hypothesis of zero correlation. In fact, it is not until reaching a postulated correlation coefficient 53\% between occupations and 77\% within occupations that we fail to reject the null hypothesis.\footnote{These results indicate that our framework is relatively more capable of ranking attributes for a specific occupation.} Together, the validation exercises suggest a significant non-zero mean correlation between the O*NET scores and our estimates.

\begin{table}[!t]
\caption{Correlation between O*NET measure and our generalized measure}
\begin{subtable}{.475\textwidth}
	\caption{\textit{Between} occupations}
	\begin{adjustbox}{max totalsize = {\textwidth}{0.9\textheight}, center}
		\begin{tabular}{rrr}
\toprule
 $\rho=\rho_{0}$ &  $t$-stat &  $p$-value \\
\midrule
           0.000 &    38.267 &      0.000 \\
           0.530 &     1.925 &      0.055 \\
\bottomrule
\end{tabular}

	\end{adjustbox}\VerticalSpaceFloat
	\label{Tbl:correlation_between_occupations}
\end{subtable}\hfill
\begin{subtable}{.475\textwidth}
	\caption{\textit{Within} occupations}
	\begin{adjustbox}{max totalsize = {\textwidth}{0.9\textheight}, center}
		\begin{tabular}{rrr}
\toprule
 $\rho=\rho_{0}$ &  $t$-stat &  $p$-value \\
\midrule
           0.000 &   265.095 &      0.000 \\
           0.770 &    -0.173 &      0.863 \\
\bottomrule
\end{tabular}

	\end{adjustbox}\VerticalSpaceFloat
	\label{Tbl:correlation_within_occupations}
\end{subtable}
\label{Tbl:correlation_between_measures}
{\footnotesize
	\textit{Notes:} This table shows the results from several $t$-tests that examine various null hypotheses of the form $\mathcal{H}_{0}:\rho=\rho_{0}$ on the correlation between the O*NET measures and our generalized measures. For each table, the first row covers the null hypothesis that the correlation is not significantly different from zero ($\rho_{0}=0$), whereas the second row highlights the first null hypothesis that leads to a $p$-value above 5\% in the testing sequence $\rho_{0}=0.01,\ldots,0.99$.  Likewise for each table, the first column is the postulated null hypothesis ($\rho$), the second column is the $t$-statistics, and the third column is the corresponding $p$-value. Table \ref{Tbl:correlation_between_occupations} considers the correlation between occupations, whereas Table \ref{Tbl:correlation_within_occupations} considers the correlation within occupations.\par}
\end{table}

\paragraph*{Overall explainability}
As a second validation exercise, we run various regressions of the O*NET scores on our estimates (and a constant), using the entire set of 213,012 attribute estimates.  The specifications we consider are the eight combinations that may be generated using occupation, attribute, and category dummies, respectively. That is, the baseline specification includes no control dummies, whereas the full specification includes all three sets of dummies.  In Table \ref{Tbl:reg_onet_on_measure}, we report the results from all eight regressions. The first row shows the coefficient on our measure, the standard error in parentheses, and the $t$-statistic in brackets.  Rows 2-4 details the specification and the last two rows show the adjusted $R^2$ and the number of observations, respectively. 

\VerticalSpace
From Table \ref{Tbl:reg_onet_on_measure}, the coefficient under scrutiny is highly significant and strongly robust across all specifications, with $t$-statistics above 300 in all specifications. In addition, the adjusted $R^2$ in the baseline specification reaches 59\%. This supports the consistency,  reliability, and validity of our proposed framework. 

\begin{table}[!t]
\caption{Explainability by regression}
	\begin{adjustbox}{max totalsize = {\textwidth}{0.9\textheight}, center}
		\begin{tabular}{lllllllll}
\toprule
{} &                    (1) &                    (2) &                    (3) &                    (4) &                    (5) &                    (6) &                    (7) &                    (8) \\
\midrule
occ2vec predicted O*NET     &         $1.1838^{***}$ &         $1.1863^{***}$ &         $1.2775^{***}$ &         $1.2149^{***}$ &         $1.2911^{***}$ &         $1.2189^{***}$ &         $1.2775^{***}$ &         $1.2911^{***}$ \\
                            &  $\left(0.0018\right)$ &  $\left(0.0018\right)$ &  $\left(0.0042\right)$ &  $\left(0.0023\right)$ &  $\left(0.0040\right)$ &  $\left(0.0022\right)$ &  $\left(0.0042\right)$ &  $\left(0.0040\right)$ \\
                            &  $\left[649.69\right]$ &  $\left[677.26\right]$ &  $\left[303.93\right]$ &  $\left[529.50\right]$ &  $\left[321.17\right]$ &  $\left[551.82\right]$ &  $\left[303.90\right]$ &  $\left[321.22\right]$ \\
Job dummies included        &                  False &                   True &                  False &                  False &                   True &                   True &                  False &                   True \\
Descriptor dummies included &                  False &                  False &                   True &                  False &                   True &                  False &                   True &                   True \\
Category dummies included   &                  False &                  False &                  False &                   True &                  False &                   True &                   True &                   True \\
$\text{Adj. }R^{2}$         &                  0.590 &                  0.629 &                  0.721 &                  0.601 &                  0.760 &                  0.640 &                  0.721 &                  0.760 \\
N                           &                213,012 &                213,012 &                213,012 &                213,012 &                213,012 &                213,012 &                213,012 &                213,012 \\
\bottomrule
\end{tabular}

	\end{adjustbox}\VerticalSpaceFloat
\label{Tbl:reg_onet_on_measure}
{\footnotesize
\textit{Notes:} This table shows the results from regressing the O*NET scores on our estimates using eight specifications that each represent a column.  Column (1) represents the baseline specification with no control, whereas column (8) represents the full specification with occupation, descriptor, and category dummies included, respectively. Columns (2)-(7) represent the specifications in between the baseline and full specification as detailed in rows 2-4.  Coefficients estimates are shown in the first row alongside standard errors in parentheses and $t$-statistics in brackets.  Superscripts ***, **, and * indicate statistical significance based on a (two-sided) $t$-test using heteroskedasticity-robust standard errors at significance levels 1\%, 5\%, and 10\%, respectively. \par}
\end{table} 
\section{Applications}\label{sec:Application}
In this section, we estimate the occupational degree of both some well-known characteristics (task measures and AI measures, respectively),  and some novel characteristics (charisma and EQ, respectively). We show how the \OccToVecS framework ranks occupations according to the characteristics and how our estimates correlate with various occupational statistics, e.g., educational requirements and wage. All occupational statistics are sourced from the U.S.  Bureau of Labor Statistics. For the well-known characteristics, we also assess the similarity of our estimates to those already established in the literature. 

\subsection{Well-known characteristics}

\subsubsection{The occupational degree of task measures}
We revisit the seminal work of \cite{Autor2003,Autor2006,Acemoglu2011,Autor2013skill,Autor2013} to study tasks measures, namely abstract, manual, and routine tasks. In Table \ref{Tbl:task_measures} in Appendix \ref{App:tasks}, we outline the individual O*NET attributes that comprise the tasks measures and refer to \cite{Acemoglu2011} for details on precisely how to construct them. We will mainly focus on the degree of routine tasks and refer to Appendix \ref{App:tasks} for similar analyses on the degree of abstract and manual tasks, respectively.

\paragraph*{Routine tasks}
We start by defining routine tasks in Definition \ref{Def:routine_taks}, relying heavily on \cite{Acemoglu2011}.\footnote{We thank Daron Acemoglu for providing feedback on this definition.} Recall how we feed this definition into our NLP method, which returns an embedding in the same vector space as the occupations. We then measure the occupational degree of routine tasks by the cosine similarity between each occupational embedding and the embedding of routine tasks.

\begin{definition}[Routine tasks]\label{Def:routine_taks}
\textquote{Routine tasks are cognitive or physical tasks, which follow closely prescribed sets of precise rules and well-defined procedures and are executed in a well-controlled environment. These routine tasks are increasingly codified in computer software and performed by machines or sent electronically to foreign work-sites to be performed by comparatively low-wage workers. Routine tasks are characteristic of many middle-skilled cognitive and production activities, such as bookkeeping, clerical work and repetitive production tasks.}
\end{definition}

In Table \ref{Tbl:top_and_bottom10_routine_task}, we list the top and bottom 10 occupations on the degree of routine tasks. The results are as expected and the NLP algorithm even picks up the specific examples in Definition \ref{Def:routine_taks} such as clerical work (e.g., rank 2) and bookkeeping (e.g., rank 9). All top 10 occupations are truly characterized by accomplishing codifiable tasks that can be specified as a series of instructions to be executed by a machine. In stark contrast, the tasks of choreographers or psychiatrists are by no means codifiable.

\begin{table}[!t]
\caption{Top and bottom 10 occupations on degree of routine tasks}
	\begin{adjustbox}{max totalsize = {\textwidth}{0.9\textheight}, center}
		\begin{tabular}{lll}
\toprule
{} &                                             Top 10 &                                          Bottom 10 \\
Rank &                                                    &                                                    \\
\midrule
1    &                        Word Processors and Typists &                                     Choreographers \\
2    &                             Office Clerks, General &                         Sports Medicine Physicians \\
3    &                                  Data Entry Keyers &                                   Music Therapists \\
4    &  Secretaries and Administrative Assistants, Exc... &                 Environmental Restoration Planners \\
5    &                    Maids and Housekeeping Cleaners &                                      Psychiatrists \\
6    &          Office Machine Operators, Except Computer &                      Chief Sustainability Officers \\
7    &                     Payroll and Timekeeping Clerks &                Social Work Teachers, Postsecondary \\
8    &  Cleaning, Washing, and Metal Pickling Equipmen... &  Substance Abuse and Behavioral Disorder Counse... \\
9    &       Bookkeeping, Accounting, and Auditing Clerks &           Fire-Prevention and Protection Engineers \\
10   &                     Machine Feeders and Offbearers &  Agents and Business Managers of Artists, Perfo... \\
\bottomrule
\end{tabular}

	\end{adjustbox}\VerticalSpaceFloat
\label{Tbl:top_and_bottom10_routine_task}
{\footnotesize
\textit{Notes:} This table shows the top and bottom 10 occupations on degree of routine tasks estimated by the \OccToVecS framework as the cosine similarity between occupation embeddings and the embedding of routine tasks.\par}
\end{table}

\begin{figure}[!t]
\begin{adjustbox}{max totalsize = {\textwidth}{\SingleFigMaxHeight\textheight}, center}
\includegraphics[width = \textwidth]{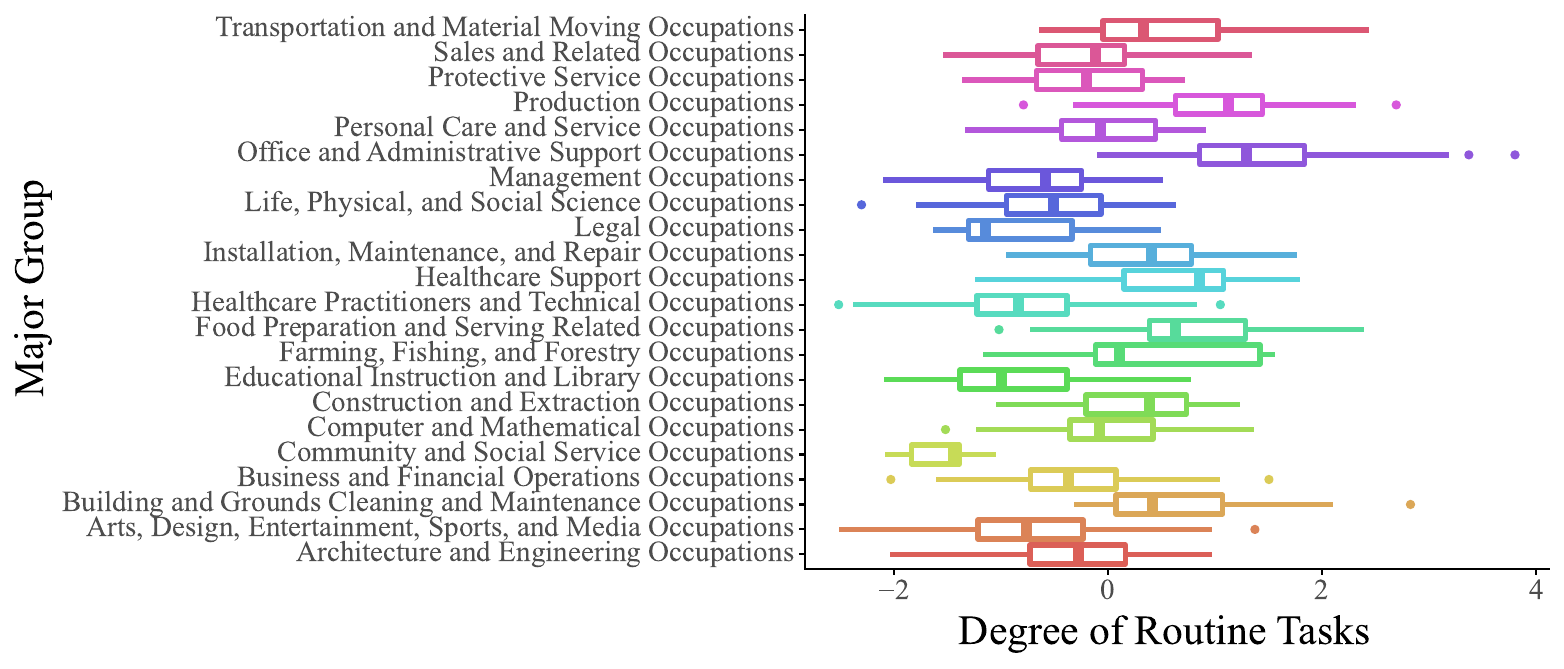}
\end{adjustbox}	
\caption{Routine tasks by major occupational group}
\label{Fig:routine_tasks_on_major_group_title}
\VerticalSpaceFloat
{\footnotesize
\textit{Notes:} This figure shows a boxplot of the occupational degree of routine tasks by major occupational groups. \par}
\end{figure}

\begin{figure}[!t]
\begin{adjustbox}{max totalsize = {\textwidth}{\SingleFigMaxHeight\textheight}, center}
\includegraphics[width = \textwidth]{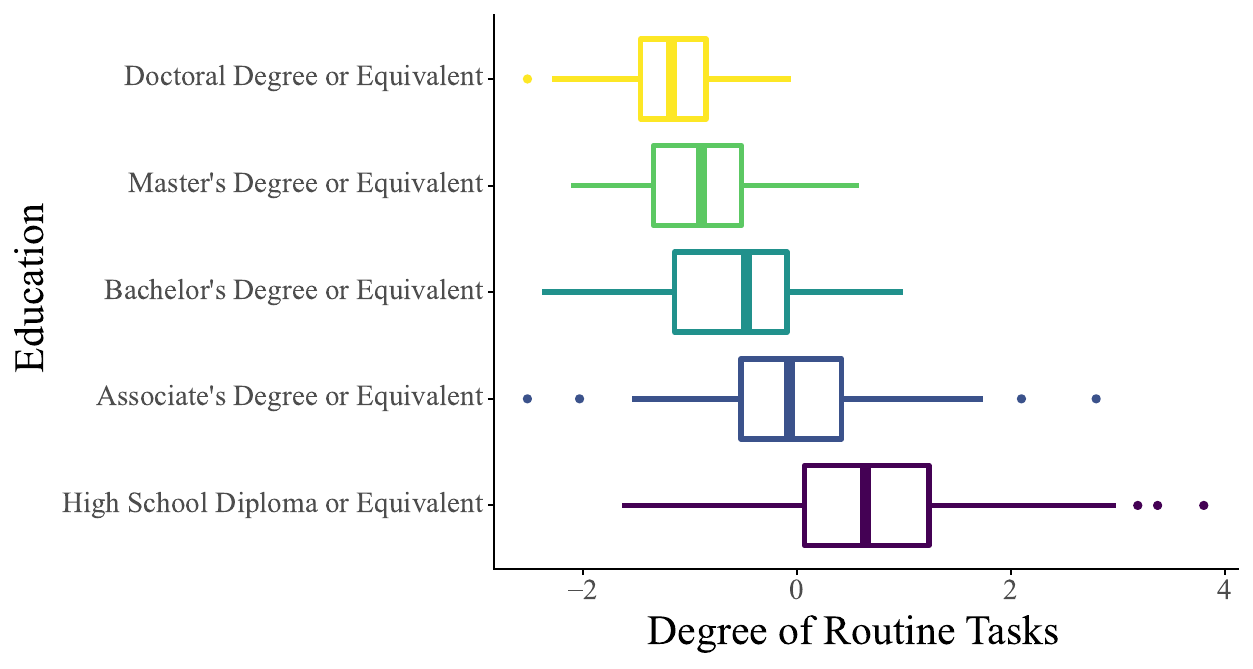}
\end{adjustbox}	
\caption{Routine tasks by education}
\label{Fig:routine_tasks_on_education}
\VerticalSpaceFloat
{\footnotesize
\textit{Notes:} This figure shows a boxplot of the occupational degree of routine tasks by educational requirement. \par}
\end{figure}

\VerticalSpace
Given the ability of the framework to separate occupations by major occupational groups and educational requirements, we show the degree of routine tasks by these two attributes in Figure \ref{Fig:routine_tasks_on_major_group_title} and \ref{Fig:routine_tasks_on_education}, respectively.  Figure \ref{Fig:routine_tasks_on_major_group_title} confirms that occupations that accomplish a higher degree of routine tasks are to be found in office and administrative support occupations, where the opposite can be found in community and social service occupations as well as educational instruction and library occupations. Figure \ref{Fig:routine_tasks_on_education} further shows that routine tasks are characteristic of occupations that require low-to-medium educational attainment,  where occupations that require bachelor's, master's or doctoral degrees rarely involve routine tasks. In fact, the occupational degree of charisma and the educational requirement are inversely, monotonically related.

\begin{figure}[!t]
		\begin{adjustbox}{max totalsize = {\textwidth}{\SingleFigMaxHeight\textheight}, center}
			\includegraphics[width = \textwidth]{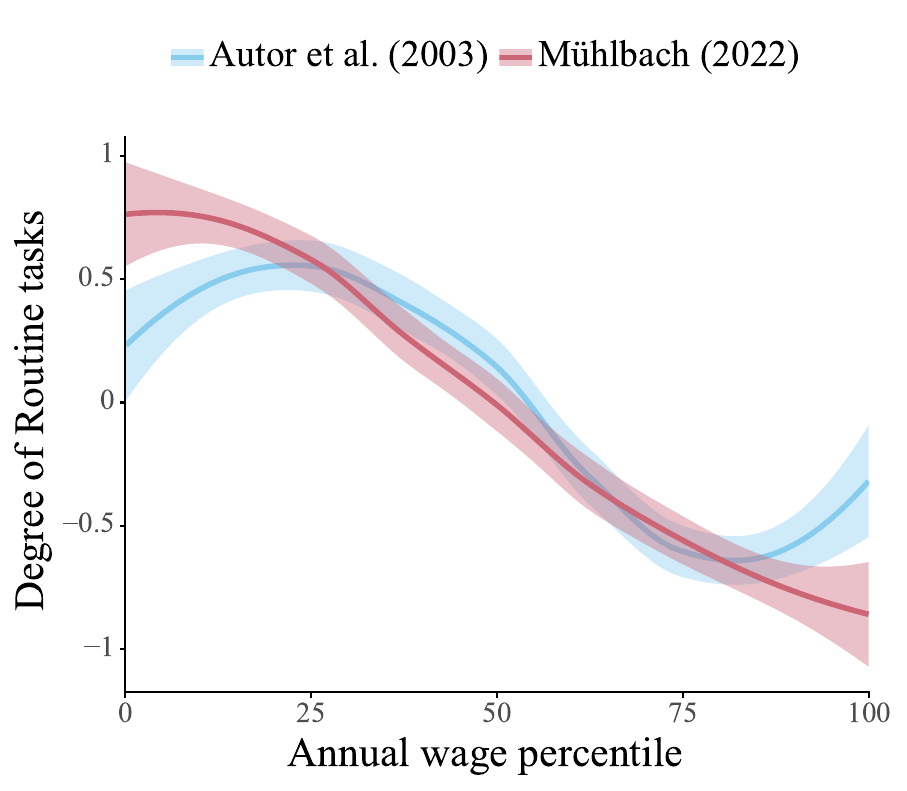}
		\end{adjustbox}	
	\caption{Comparing estimates of routine tasks on wage}
	\label{Fig:comparing_routine_tasks_to_expert_measure}
\VerticalSpaceFloat
{\footnotesize
\textit{Notes:} This figure shows a smoothed polynomial regression of both standardized measures of routine tasks in each 6-digit SOC occupation against its rank in the wage distribution. \par}
\end{figure}

We next highlight how our estimates largely coincide with the original measures by \cite{Autor2003}. Essentially, we estimate a smoothed polynomial regression of both measures of routine tasks for each occupation against its rank in the wage distribution. The result is shown in Figure \ref{Fig:comparing_routine_tasks_to_expert_measure}. Overall, the two approaches agree nearly perfectly on the relationship between the degree of routine tasks and the wage rank; occupations in the upper part of the wage distribution tend to be characterized by smaller amounts of routine tasks, whereas the opposite holds for the lower part of the wage distribution where mainly occupations with many routine tasks reside.  A trend that has been evidenced by many (see, e.g., \cite{Acemoglu2011}).

\VerticalSpace
As a final comparison, we compute the Spearman correlation coefficients between our estimates of abstract, manual, and routine tasks and the original ones.  The correlation coefficients are 0.40, 0.84, and 0.64, respectively. Thus, the degree of association between the two approaches appears strong and significant, particularly for manual and routine tasks.

\subsubsection{The occupational degree to artificial intelligence}
Turning to the occupational degree of AI, we use the Definitions \ref{Def:artificial_general_intelligence}--\ref{Def:prescriptive_analytics} presented in Appendix \ref{App:ai}. Similar, all figures and tables for this application is deferred to Appendix \ref{App:ai}. We start by considering the top and bottom 10 occupations on degree of AI from Table \ref{Tbl:top_and_bottom10_artificial_intelligence}. In top 10, we find e.g.,  business intelligence analysts and statisticians, which is aligned with our expectation as these occupations involve tasks that require various forms of e.g., predictive and prescriptive analytics, which are integral parts of AI. In the bottom 10, we find e.g., manicurists, pedicurists, and stonemasons.

\VerticalSpace
Considering the degree of AI by major occupational groups and educational requirement, we show Figure \ref{Fig:artificial_intelligence_on_major_group_title} and Figure \ref{Fig:artificial_intelligence_on_education}, respectively. As expected, the top major occupational groups computer and mathematical occupations, which coincides with the top occupations from Table \ref{Tbl:top_and_bottom10_artificial_intelligence}.  The educational requirements tend to be higher for occupations with high degrees of AI, which makes sense as AI is difficult to skill to acquire, but the association is not monotonically increasing in education. In fact, occupations requiring bachelor's or master's degrees tend to have a higher exposure to AI than occupations that require doctoral degrees.  

\VerticalSpace
Next, we compare our estimates of occupational degree of AI to the already established measures, namely those of \cite{Felten2018},  \cite{Brynjolfsson2018a},  and \cite{Webb2019}.  In Figure \ref{Fig:comparing_concepts_artificial_intelligence_on_stat_annual_wage}, we repeat the previous analysis and estimate a smoothed polynomial of the measures for each occupation against its rank in the wage distribution.  Our measure of the occupational degree of AI is to be found in the middle between the measures of \cite{Felten2018} and \cite{Webb2019}, meaning that its relationship to occupational wage and employment growth is a mix of those of these well-established measures.  Interestingly,  our measure of occupational AI tends to agree relatively more with the one of \cite{Felten2018} for the high-wage occupations and agree relatively more with the one of \cite{Webb2019} for the low-wage occupations. In fact, the Spearman correlation coefficient between our measure and those two is 0.35 and 0.2,  respectively, as shown in Table \ref{Tbl:correlation_coefficients_artificial_intelligence}.  Specifically, occupations receiving higher wages tend to have a higher degree of AI compared to the occupations that receive less.  The differences between the AI--wage profiles shown in Figure \ref{Fig:comparing_concepts_artificial_intelligence_on_stat_annual_wage}, however,  suggest that the three established measures are picking up different aspects of AI. A fact that is already found by \cite{Acemoglu2022ai}, who also find that the measures of \cite{Felten2018} and \cite{Webb2019} tend to coincide and be different than the one of \cite{Brynjolfsson2018a}.

\subsection{Novel characteristics}
The second part of this section considers two novel characteristics of occupations, namely the degree of charisma and EQ.  To economize on space, all figures and tables for the EQ application are deferred to Appendix \ref{App:eq}.  Definitions are likewise found in Appendix \ref{App:charisma} and \ref{App:eq}, respectively.

\subsubsection{The occupational extent of charisma}
The first systematic treatment of charisma is due to \cite{Weber1947} and popularized by \cite{Dow1969}. Charisma is at the very center of effective leadership \citep{Avolio2013} and is continuously being studied (see, e.g., \cite{Hippel2016}). It has been recognized as one of the main explanations for why certain leaders, i.e., charismatic leaders, develop emotional attachment with followers and other leaders that eventually foster performance that surpass expectations. Likewise, charisma is associated with being considerate, inspirational, visionary and intellectually stimulating (for a review, see, e.g., \cite{Spencer_1973} or \cite{Turner_2003}). One definition of charisma follows from Definition \ref{Def:charisma_apa}, which is part of the definitions we use to construct the charisma embedding.\footnote{The other definitions that are included in the construction of the embedding of charisma follow from Definitions \ref{Def:charisma_psyc}--\ref{Def:charisma_wiki} in Appendix \ref{App:charisma}.}

\begin{definition}[Charisma]\label{Def:charisma_apa}
\textquote{The special quality of personality that enables an individual to attract and gain the confidence of large numbers of people. It is exemplified in outstanding political, social, and religious leaders.} \citep{APA_charisma}
\end{definition}

\begin{figure}[!t]
\begin{adjustbox}{max totalsize = {\textwidth}{\SingleFigMaxHeight\textheight}, center}
\includegraphics[width = \textwidth]{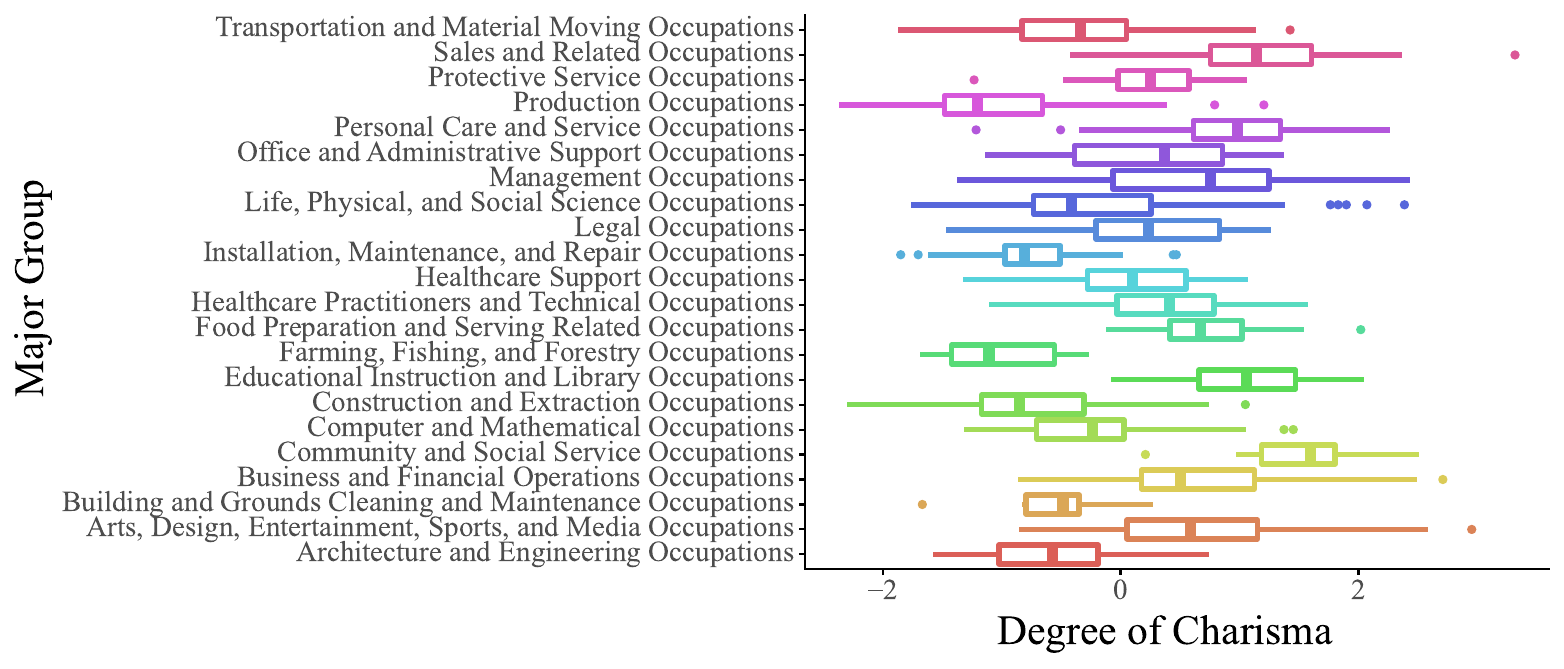}
\end{adjustbox}	
\caption{Charisma by major occupational group}
\label{Fig:charisma_on_major_group_title}
\VerticalSpaceFloat
{\footnotesize
\textit{Notes:} This figure shows a boxplot of the occupational degree of charisma by major occupational groups. \par}
\end{figure}

We begin our analysis by highlighting the top and bottom 10 occupations on the degree of charisma in Table \ref{Tbl:top_and_bottom10_charisma}.  Public Relations specialists, actors,  and fundraisers are among the top occupations that associate with charisma,  because those occupations require both the ability to catch the attention of the audience and the knowledge of group behavior and dynamics, societal trends and influences.  Similar for school psychologists who must be knowledgeable of human behavior and be capable of assessing individual differences in personality and interests.  An interesting yet obvious finding is that directors within religious activities and education are among the top occupations as these directors must be able to speak to the public and gain followers by promoting the religious education or activities and teaching their religion's doctrines.  Generally,  occupations within community and social services as well as sales score high on charisma. On the other side of the spectrum, we mostly find occupations within production, construction, farming, fishing,  and forestry. These occupations have less of a need to be able to inspire and motivate large numbers of people.  This finding is confirmed by Figure \ref{Fig:charisma_on_major_group_title}, showing the occupational degree of charisma by major occupational groups. 

\begin{table}[!t]
\caption{Top and bottom 10 occupations on degree of charisma}
	\begin{adjustbox}{max totalsize = {\textwidth}{0.9\textheight}, center}
		\begin{tabular}{lll}
\toprule
{} &                                         Top 10 &                                          Bottom 10 \\
Rank &                                                &                                                    \\
\midrule
1    &                                  Telemarketers &                    Biofuels Processing Technicians \\
2    &                   Public Relations Specialists &  Woodworking Machine Setters, Operators, and Te... \\
3    &                                    Fundraisers &                                     Fence Erectors \\
4    &                                         Actors &                        Floor Sanders and Finishers \\
5    &  Directors, Religious Activities and Education &  Separating, Filtering, Clarifying, Precipitati... \\
6    &                   Search Marketing Strategists &  Textile Winding, Twisting, and Drawing Out Mac... \\
7    &                       Human Resources Managers &  Extruding, Forming, Pressing, and Compacting M... \\
8    &                           School Psychologists &  Extruding and Forming Machine Setters, Operato... \\
9    &                            Writers and Authors &                             Potters, Manufacturing \\
10   &                       Advertising Sales Agents &  Paving, Surfacing, and Tamping Equipment Opera... \\
\bottomrule
\end{tabular}

	\end{adjustbox}\VerticalSpaceFloat
\label{Tbl:top_and_bottom10_charisma}
{\footnotesize
\textit{Notes:} This table shows the top and bottom 10 occupations on degree of charisma estimated by the \OccToVecS framework.\par}
\end{table}

\begin{figure}[!t]
\begin{adjustbox}{max totalsize = {\textwidth}{\SingleFigMaxHeight\textheight}, center}
\includegraphics[width = \textwidth]{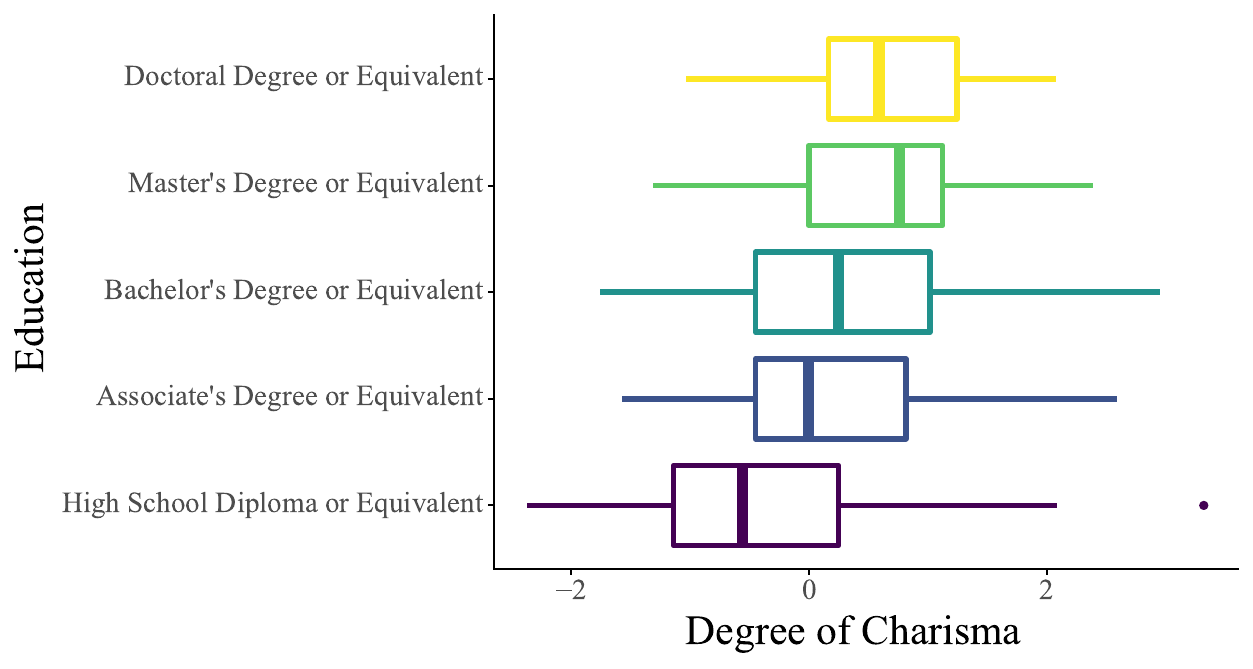}
\end{adjustbox}	
\caption{Charisma by education}
\label{Fig:charisma_on_education}
\VerticalSpaceFloat
{\footnotesize
\textit{Notes:} This figure shows a boxplot of the occupational degree of charisma by educational requirement. \par}
\end{figure}

Interestingly,  an almost monotonic pattern appears between education requirements and degree of charisma as shown in Figure \ref{Fig:charisma_on_education}. The educational requirements that score the lowest on charisma is high school diploma or equivalent, whereas occupations with educational requirements of bachelor's degrees or higher are increasingly associated with charisma.

\begin{figure}[!t]
	\begin{adjustbox}{max totalsize = {\textwidth}{\SingleFigMaxHeight\textheight}, center}
			\includegraphics[width = \textwidth]{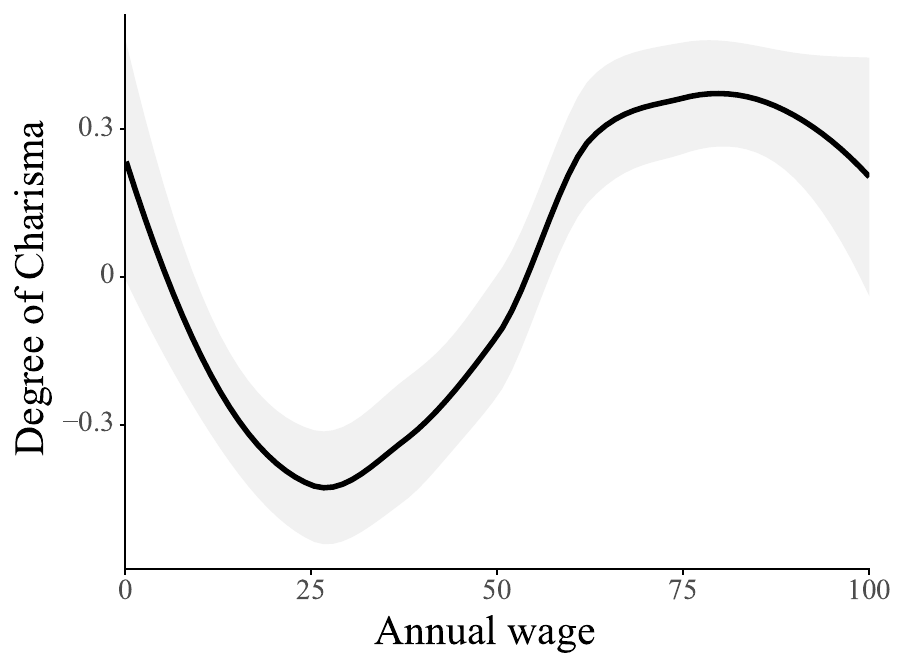}
		\end{adjustbox}	
	\caption{Charisma by occupation wages}
	\label{Fig:charisma_on_annual_wage_percentile}
\VerticalSpaceFloat
{\footnotesize
\textit{Notes:} This figure shows a smoothed polynomial regression of our standardized measure of charisma in each 6-digit SOC occupation against its rank in the wage distribution.\par}
\end{figure}

\VerticalSpace
Last, we examine how charisma is connected to wage in Figure \ref{Fig:charisma_on_annual_wage_percentile} by the means of a smoothed polynomial regression.  Figure \ref{Fig:charisma_on_annual_wage_percentile} indicates that there is a non-monotonic relationship between charisma and wage.  Essentially, in the lowest part of the wage distribution, occupations tend to have high degrees of charisma. These occupations are, e.g., actors (presumably non-Hollywood actors), teachers, and social workers.  In contrast,  the occupations with the lowest score on charisma are on average found between the first and second quartile of the wage distribution, whereas occupations in the upper part of the wage distribution can be characterized by high degrees of charisma. The latter would be the occupations that require advanced degrees and leadership capabilities. 

\subsubsection{The occupational extent of emotional intelligence}
Although the term gained popularity by \cite{Goleman1995}, the concept of EQ has been well-studied before starting with \cite{Maslow1950,Beldoch1964,Leuner1966}. Generally speaking, EQ refers to the ability to intelligently perceive, understand,  and manage emotions, and people with high EQ use this ability to guide behavior. The occupational degree of EQ is particularly interesting as EQ has been shown to correlate positively with many desirable outcomes. In summary, a recent review by \cite{Mayer2008} documents that higher EQ correlates with better social relations,  better perceptions by others, better academic achievement, and better general well-being, e.g., higher life satisfaction. 

\VerticalSpace
We begin the analysis with Table \ref{Tbl:top_and_bottom10_emotional_intelligence}, tabulating the top and bottom ten occupations on EQ, and Figure \ref{Fig:emotional_intelligence_on_major_group_title} showing EQ by occupational groups.  Many top 10 occupations are either therapists,  counselors,  or psychologists, and these occupations truly require the ability to carry out reasoning about emotions to support others in various needs.  Contrary in the bottom 10,  these occupations, e.g., brickmasons, riggers, or tapers,  require less emotional knowledge and more arm-hand steadiness and manual dexterity.  This applies more generally to occupations within construction, maintenance, installation, and repair.

\VerticalSpace
An interesting tail effect is found when considering EQ by educational requirements in Figure \ref{Fig:emotional_intelligence_on_education}.  We generally find a monotonic relationship from low education and low EQ to high education and high EQ, but this monotonicity stops once we consider occupations that on average require a doctorate.  This indicates that occupations that require doctorates rather than people with a master's degree need very specialized employees that may not have to engage too often with a larger group of coworkers rather than focusing deeply on very complex problems.

\VerticalSpace
Last, we consider the usual smoothed regression against ranks in the wage distribution in Figure \ref{Fig:emotional_intelligence_on_annual_wage_percentile}. The EQ-wage relationship from Figure \ref{Fig:emotional_intelligence_on_annual_wage_percentile} is slightly concave for low-wage occupations and slightly convex and tapering off for high-wage occupations, but generally high-wage occupations are associated with high levels of EQ.  That EQ should be positively correlated with wage at the individual level is also found by \cite{Rode_2017},  however,  the full distribution tells us that tail effects do exists.  The highly nonlinear relationship between occupational charisma and wage in Figure \ref{Fig:charisma_on_annual_wage_percentile} cannot, however,  is not found EQ. This tentatively suggests that EQ is on average a more difficult characteristic to achieve compared to charisma in terms of wage returns. However, we emphasize that these relationships are mere correlations and we do not attempt to draw any causal inference.  
\section{Conclusion}\label{sec:Conclusion}
It is inherently interesting to study occupational characteristics and how they associate with descriptive statistics of the labor market, e.g., wage, educational attainment, labor force participation rates, etc.  These studies inform policy-makers on how to optimally design education and labor markets. But estimating the occupational degree of certain characteristics is challenging for many reasons. Most importantly,  it may be severely difficult to validate and verify novel measures of occupational characteristics once constructed.  This implicitly leaves too much room for research creativity with regards to the selection of occupational features that enter the composition of new ones.

\VerticalSpace
We propose \OccToVecS as a fully data-driven and principal approach to quantifying occupations, which enable the measurement of any occupational characteristic of interest in a transparent and verifiable way.  Using NLP, we embed more than 17,000 occupation-specific descriptors sourced from the O*NET as vectors and combine them into unique occupation vectors of high dimensions. Using an objective and reliable definition of a given target characteristic of interest, we also embed this into a high-dimensional vectors and measure the occupational degree of the given characteristic as the cosine similarity between the vectors.

\VerticalSpace
Using the actual scores from O*NET on 244 attributes common to the universe of occupations as the ground truth, we extensively validate our approach by comparing our estimates to those scores. We find that our estimates explain almost 60\% of the variation in the original scores and, joint with other validation exercises, we take this as evidence that our framework is capable of producing high-quality occupation vectors that accurately capture detailed aspects of occupations. 

\VerticalSpace
We apply \OccToVecS to four applications, where we study the occupational degree of two known characteristics as well as two novel. As known, or previously studied, characteristics, we consider the popular task measures (abstract, manual, and routine tasks), and exposure to AI. Our estimates of the task measures largely match the original ones. For instance, we also find that occupations involving many routine or manual tasks tend to appear in the bottom of the wage distribution. The opposite holds for abstract tasks. Regarding exposure to AI, our measure also broadly agrees with the already-proposed measures and can in fact be seen as a mix of the two most widely used ones. As novel characteristics, we consider the occupational degree of charisma and EQ. We find that occupations that score high on these attributes tend to be found within community and social service occupations,  arts, entertainment, sports, and media occupations, and educational instruction occupations. High scores occur most often for occupations that require advanced degrees and in the top of the wage distribution.  The most striking difference in found for occupations in the bottom of the wage distribution, where scores tend to be high on charisma but not on EQ. This suggests that wage returns to charisma and EQ could be very different.

\VerticalSpace
In summary, this paper proposes a data-driven and universal approach to representing occupations as mathematical objects. This has many interesting economic applications, and particularly, we demonstrate how one may advance from purely qualitative definitions to quantitative scores at the occupational level.  The \OccToVecS framework, therefore, opens many doors for future analyses that study how novel occupational characteristics that have previously been inaccessible to researchers matter for labor market outcomes. 
\clearpage
\phantomsection
\addcontentsline{toc}{section}{References}
\bibliographystyle{ecta}

\newpage

\newpage 
\appendix
\renewcommand{\thesection}{\Alph{section}}
\numberwithin{equation}{section}
\numberwithin{figure}{section}
\numberwithin{table}{section}

\FloatBarrier
\clearpage
\section{Data appendix}\label{App:data}
In this appendix, we provide additional details to those given in Section \ref{sec:Data} on the data used. This appendix is meant as a supplement and is not self-contained without the main text. 

\begin{table}[!ht]
\caption{O*NET scales}
	\begin{adjustbox}{max totalsize = {\textwidth}{0.9\textheight}, center}
\begin{tabular}{p{0.17\linewidth}p{0.3\linewidth}p{0.7\linewidth}}
\toprule 
Scale & Category & Definition\tabularnewline
\midrule
Importance & Tasks, Knowledge, Skills, Abilities, Work Activities, and Work Styles & Rating indicates the degree of importance a particular descriptor
is to the occupation.\tabularnewline
Level & Knowledge, Skills, Abilities, and Work Activities & Rating indicates the degree to which a particular descriptor is required
or needed to perform the occupation.\tabularnewline
Relevance & Tasks & Rating indicates the degree to which a particular task is relevant
to perform the occupation.\tabularnewline
Frequency & Tasks & Rating indicates how frequent a particular task is performed within
a given time period.\tabularnewline
Occupational Interest  & Interests & Rating indicates the degree to which a particular interest \citep{Holland1985} matches the occupation.\tabularnewline
Extent & Work Values & Rating indicates the degree to which a particular work value affects
the nature of the occupation.\tabularnewline
Context & Work Context & Rating indicates the degree to which a particular work context influences
the nature of the occupation.\tabularnewline
\bottomrule
\end{tabular}
	\end{adjustbox}\VerticalSpaceFloat
\label{Tbl:onet_scales}
{\footnotesize
\textit{Notes:} This table shows scales associated with each descriptor within categories.  All scales are standardized to a scale ranging from 0 to 1.  See O*NET site on scales, ratings, and standardized scores for more information (\href{https://www.onetonline.org/help/online/scales}{here}).\par}
\end{table}
 
\FloatBarrier
\clearpage
\section{Method appendix}\label{App:method}
In this appendix, we provide additional details to those given in Section \ref{sec:Methodology} on the methods applied. This appendix is meant as a supplement and is not self-contained without the main text. 

\paragraph*{BERT---Deep Bidirectional Transformers} BERT stands for Bidirectional Encoder Representations from Transformers and is a revolutionary technique to learning general-purpose language representations, using a deep bidirectional Transformer encoder based on the original Transformer encoder-decoder from \cite{Vaswani2017}.  The bidirectional aspect of BERT means that it simultaneously learns a language representation from left to right \textit{and} from right to left compared to either a left-to-right model or the shallow concatenation of a left-to-right and a right-to-left model. Upon release, BERT achieved state-of-the-art results on a range of NLP tasks. Because the use of Transformers has become common, we will omit an exhaustive description of the attention architecture and refer readers to \cite{Vaswani2017}. 

\VerticalSpace
In general, BERT consists of two steps, namely a pre-training and a fine-tuning step.  First, the pre-training step encodes a lot of semantics and syntactic information about the language by training the model on massive amounts of unlabeled textual data drawn from the web. The objective is to detect generic linguistic patterns in the text, and it is mathematically represented as the sum of two separate objectives, that is, the Masked Language Model (MLM) objective and the Next Sentence Prediction (NSP) objective. 
The intuition behind MLM is that deep language knowledge is required to be able to filling in the blanks \textit{within} a sentence with missing words. This in return requires one to read the entire sentence and that is why the bidirectional conditioning is important. The MLM is essentially a cloze procedure \citep{Taylor_1953} that masks a random sample of tokens by replacing the given token with the $\left[MASK\right]$ token. The MLM objective is then a cross-entropy loss on predicting the masked tokens, where the cross-entropy loss for a specific masked token, $t_0$, is
\begin{align}\label{Eq:cross_entropy_loss}
	H\left(t_{0}\right)=-\sum_{c=1}^{\left|\mathcal{V}\right|}y_{t_{0},c}\log p_{t_{0},c},
\end{align}
where $y_{t_{0},c}=\mathds{1}_{\left\{ c:p_{t_{0},c}=\max_{c\in\mathcal{V}}p_{t_{0},c}\right\} =c}$ is a binary indicator that indicates if the (predicted) class $c$ is correct for token $t_0$,  and $p_{t_{0},c}$ is the (predicted) probability that token $t_0$ is of $c$. BERT uniformly selects 15\% of the input tokens and perform one of three replacements on the selected tokens; (1) 80\% are replaced with $\left[MASK\right]$,  (2) 10\% are left unchanged, and (3) 10\% are replaced with a randomly selected token from vocabulary $\mathcal{V}$. In theory, BERT performs MLM once in the beginning, although in practice, the masks are not the same for every sequence because the data are duplicated for parallel training. 

\VerticalSpace
In contrast, NSP is used for understanding the relationship \textit{between} sentences and it similarly a classification problem. But where MLM considers multiclass classification,  however, NSP uses a binary classification loss for whether one sentence is adjacent to another in the original texts.  BERT optimizes the sum of the two loss functions using ADAM \citep{Kingma2015} and the parameters can be found in Table \ref{Tbl:hyperparam} further below.

\VerticalSpace
Second,  fine-tuning is used to optimize performance for a specific downstream task, e.g., question answering, text summarization, or sentence embedding, and BERT is thus an example of transfer learning that can use that language representation learnt in the pre-training step across many specific language tasks. Fine-tuning is often performed by adding one or more layers on top of the deep network obtained under pre-training.  We follow \cite{reimers2019} and fine-tune BERT/RoBERTa to yield useful sentence embeddings by training the model on labeled sentence pairs and minimizing the mean-squared-error (MSE) loss between sentence embeddings. In principal, the pre-training step is sufficient to construct the text embeddings but \cite{reimers2019} demonstrate that these embeddings are not always meaningful. For this reason, we add the fine-tuning step of \cite{reimers2019} to construct the sentence embeddings. In addition, we implement another fine-tuning step of estimating a rotation matrix tailored to \OccToVec, which we describe in the main text.

\paragraph*{RoBERTa---A robustly optimized BERT}
\cite{Liu2019} find that the original implementation of BERT was significantly undertrained and propose RoBERTa, a robustly optimized version of BERT.  RoBERTa builds on the same Transformer model as BERT and also uses the MLM objective,  where it learns to predict intentionally hidden pieces of texts.  The architectural difference is that RoBERTa drops the NSP objective and updates the MLM strategy to be dynamic within training epochs. In addition, RoBERTa is trained on more data and modifies some hyperparameters in the training process compared to BERT, e.g.,  larger batches and step sizes, leading to 4-5 times longer training times. The advantage is a 2-20\% improvement over BERT on several benchmark tasks for NLP algorithms. In essence, RoBERTa outperforms BERT and could, at the time of release, match or exceed every NLP model published after BERT in all individual tasks on several benchmarks, including the General Language Understanding Evaluation (GLUE), the Stanford Question Answering Dataset (SQuAD), and the ReAding Comprehension from Examinations (RACE) benchmarks, which is the reason we choose RoBERTa as our preferred NLP algorithm.  We will omit repeating unnecessarily the details of the implementation of RoBERTa and refer readers to \cite{Liu2019}.

\paragraph*{Neural network architecture}
Let $L$ denote the number of layers (i.e., Transformer blocks),  let $d$ denote the hidden size (i.e., the dimensions of the embeddings), and let $A$ denote the number of self-attention heads. The model size we consider uses $L=24$, $d=1024$, and $A=16$. All hyperparameters can be found in Table \ref{Tbl:hyperparam}.  For $m=1,\ldots,N$, let $T_m$ be a piece of text (e.g., a sentence or a paragraph) from $\mathcal{T}$. Each $T_m$ is a sequence of tokens (e.g., words or subwords), $t_{m,1},\ldots,t_{m,N_{m}}$., generated by WordPiece tokenization \citep{Wu2016}.  BERT takes as input a concatenation of two sequences,  $t_{m,1},\ldots,t_{m,N_{m}}$ and $t_{m',1},\ldots,t_{m',N_{m'}}$ for $m\neq m'$.  The two sequences are delimited by special tokens, namely
\begin{align}\label{Eq:bert_input}
\textrm{tok}\left(T_{m},T_{m'}\right)=\left[CLS\right],t_{m,1},\ldots,t_{m,N_{m}},\left[SEP\right],t_{m',1},\ldots,t_{m',N_{m'}},\left[EOS\right],
\end{align}
where $\textrm{tok}\left(\cdot,\cdot\right)$ produces the special tokenization of two input texts, $CLS$ abbreviates classification, $SEP$ separation, and $EOS$ end-of-sentence.  The lengths of $T_m$ and $T_m'$ are constrained such that $N_{m}+N_{m'} < \bar{N}$, where $\bar{N}$ is fixed beforehand.  From \ref{Eq:bert_input}, BERT constructs three temporary input embeddings that, once summed, act as the input embedding to the Transformer.  First, each token $t$ in $\textrm{tok}\left(T_{m},T_{m'}\right)$ is passed through three deterministic functions, 
\begin{align}
l:\left(t,\mathcal{V}\right) &\rightarrow \left[1,\left|\mathcal{V}\right|\right], \label{Eq:lookup_token}  \\
p:\left(t,\textrm{tok}\left(T_{m},T_{m'}\right)\right) &\rightarrow \left[1,\bar{N}\right],\label{Eq:position_token} \\
s:\left(t,\textrm{tok}\left(T_{m},T_{m'}\right)\right)&\rightarrow\left[1,2\right],\label{Eq:sequence_token}
\end{align}
where $l\left(t,\mathcal{V}\right)$ is a lookup function that returns the index of token $t$ in vocabulary $\mathcal{V}$ of all tokens,  $p\left(t,\textrm{tok}\left(T_{m},T_{m'}\right)\right)$ is a position function that returns the index of token $t$ in the concatenated sequences $\textrm{tok}\left(T_{m},T_{m'}\right)$, and $s\left(t,\textrm{tok}\left(T_{m},T_{m'}\right)\right)$ is a sequence function that returns the sequence index that token $t$ belongs to in $\textrm{tok}\left(T_{m},T_{m'}\right)$. Essentially, each token is represented as three integers that determines the both position of the token in the vocabulary and the concatenated sequences,  respectively, and which of the individual sequences that the token belongs to.  Last, each integer (i.e., token, position, and sequence) is passed through an embedding layer $\phi_{h}:\mathbb{R}\rightarrow\mathbb{R}^{d}$ for $h=l,p,s$ that returns a $d$-dimensional embedding.  The three embeddings are summed to generate the final input embedding $e_{t}\equiv e\left(t;\mathcal{V},T_{m},T_{m'}\right)$. That is,
\begin{align}
e\left(t;\mathcal{V},T_{m},T_{m'}\right)=\phi_{l}\left(l\left(t,\mathcal{V}\right)\right)+\phi_{p}\left(p\left(t,\textrm{tok}\left(T_{m},T_{m'}\right)\right)\right)+\phi_{s}\left(s\left(t,\textrm{tok}\left(T_{m},T_{m'}\right)\right)\right)
\end{align}
We show the full process of turning tokens into input embeddings in Figure \ref{Fig:bert_input_embedding}. These input embeddings are then passed through the layers of the deep Transformer model. During training, BERT considers two primary objectives,  i.e., the MLM objective and the NSP objective, as described above.

\begin{figure}[!ht]
	\begin{adjustbox}{max totalsize = {\textwidth}{0.9\textheight}, center}
	$
\begin{array}{ccccccccc}
\left[CLS\right] & t_{m,1} & \ldots & t_{m,N_{m}} & \left[SEP\right] & t_{m',1} & \ldots & t_{m',N_{m}} & \left[EOS\right]\\
\\
\downarrow & \downarrow &  & \downarrow & \downarrow & \downarrow &  & \downarrow & \downarrow\\
\\
l\left(\left[CLS\right]\right) & l\left(t_{m,1}\right) & \ldots & l\left(t_{m,N_{m}}\right) & l\left(\left[SEP\right]\right) & l\left(t_{m',1}\right) & \ldots & l\left(t_{m',N_{m}}\right) & l\left(\left[EOS\right]\right)\\
\\
p\left(\left[CLS\right]\right) & p\left(t_{m,1}\right) & \ldots & p\left(t_{m,N_{m}}\right) & p\left(\left[SEP\right]\right) & p\left(t_{m',1}\right) & \ldots & p\left(t_{m',N_{m}}\right) & p\left(\left[EOS\right]\right)\\
\\
s\left(\left[CLS\right]\right) & s\left(t_{m,1}\right) & \ldots & s\left(t_{m,N_{m}}\right) & s\left(\left[SEP\right]\right) & s\left(t_{m',1}\right) & \ldots & s\left(t_{m',N_{m}}\right) & s\left(\left[EOS\right]\right)\\
\\
\downarrow & \downarrow &  & \downarrow & \downarrow & \downarrow &  & \downarrow & \downarrow\\
\\
\phi_{l}\left(l\left(\left[CLS\right]\right)\right) & \phi_{l}\left(l\left(t_{m,1}\right)\right) & \ldots & \phi_{l}\left(l\left(t_{m,N_{m}}\right)\right) & \phi_{l}\left(l\left(\left[SEP\right]\right)\right) & \phi_{l}\left(l\left(t_{m',1}\right)\right) & \ldots & \phi_{l}\left(l\left(t_{m',N_{m}}\right)\right) & \phi_{l}\left(l\left(\left[EOS\right]\right)\right)\\
+ & + &  & + & + & + &  & + & +\\
\phi_{p}\left(p\left(\left[CLS\right]\right)\right) & \phi_{p}\left(p\left(t_{m,1}\right)\right) & \ldots & \phi_{p}\left(p\left(t_{m,N_{m}}\right)\right) & \phi_{p}\left(p\left(\left[SEP\right]\right)\right) & \phi_{p}\left(p\left(t_{m',1}\right)\right) & \ldots & \phi_{p}\left(p\left(t_{m',N_{m}}\right)\right) & \phi_{p}\left(p\left(\left[EOS\right]\right)\right)\\
+ & + &  & + & + & + &  & + & +\\
\phi_{s}\left(s\left(\left[CLS\right]\right)\right) & \phi_{s}\left(s\left(t_{m,1}\right)\right) & \ldots & \phi_{s}\left(s\left(t_{m,N_{m}}\right)\right) & \phi_{s}\left(s\left(\left[SEP\right]\right)\right) & \phi_{s}\left(s\left(t_{m',1}\right)\right) & \ldots & \phi_{s}\left(s\left(t_{m',N_{m}}\right)\right) & \phi_{s}\left(s\left(\left[EOS\right]\right)\right)\\
= & = &  & = & = & = &  & = & =\\
e_{\left[CLS\right]} & e_{t_{m,1}} & \ldots & e_{t_{m,N_{m}}} & e_{\left[SEP\right]} & e_{t_{m',1}} & \ldots & e_{t_{m',N_{m}}} & e_{\left[EOS\right]}
\end{array}
$
\end{adjustbox}
\caption{BERT input embedding}
\label{Fig:bert_input_embedding}
\VerticalSpaceFloat
{\footnotesize
\textit{Notes:} This figure shows how BERT translates tokens into an input embedding by summing embeddings of the integers that describe both the token's position in the vocabulary and the concatenated sequences and which of the sequences the token belongs to. \par}
\end{figure}

\VerticalSpace
In our framework, the object of interest is the sentence or paragraph embedding of $T_m$, and generally, three strategies exist to compute these embeddings once the final word embeddings have been estimated; \textit{mean-pooling} that uses the average of all the word embeddings from $T_m$,  \textit{max-pooling} that uses the maximum of all the word embeddings from $T_m$, or \textit{first-pooling} that uses the first of all the word embeddings from $T_m$ (i.e., the $\left[CLS\right]$ token). In theory, embeddings from any of the $L$ layers can be used as word embeddings, but it is common practice to average over a range of the last $L$ layers, e.g., the last four layers. We use \textit{mean-pooling} over the last four layers to construct our initial sentence embeddings, and then we add a pooling operation to the output of BERT/RoBERTa to derive the sentence embeddings, which is proposed \cite{reimers2019}. This is essentially a fine-tuning layer such that BERT/RoBERTa generates better sentence embeddings.  For more details, we refer interested readers to \cite{reimers2019}. The hyperparameters used for pre-training follow from Table \ref{Tbl:hyperparam}.

\begin{table}[!ht]
\caption{Hyperparameters for pre-training BERT and RoBERTa}
	\begin{adjustbox}{max totalsize = {\textwidth}{0.9\textheight}, center}
		\begin{tabular}{lrr}
\toprule 
 & BERT & RoBERTa\tabularnewline
\midrule
Number of layers, $L$ & 24 & 24\tabularnewline
Hidden size, $d$ & 1024 & 1024\tabularnewline
Feed-forward/filter size & 4096 & 4096\tabularnewline
Attention heads & 16 & 16\tabularnewline
Dropout & 0.1 & 0.1\tabularnewline
Warm-up steps & 10,000 & 30,000\tabularnewline
Weight decay, $L_{2}$ & 0.01 & 0.01\tabularnewline
Peak learning rate & 0.0001 & 0.0004\tabularnewline
Learning rate decay & Linear & Linear\tabularnewline
Gradient clipping & 0.0 & 0.0\tabularnewline
Maximum tokens, $N_{0}$ & 512 & 512\tabularnewline
Minibatch size, $B$ & 256 & 8,000\tabularnewline
Max steps, $S$ & 1,000,000 & 500,000\tabularnewline
Activation function & GELU & GELU\tabularnewline
Optimizer & Adam & Adam\tabularnewline
Adam, $\beta_{1}$ & 0.9 & 0.9\tabularnewline
Adam, $\beta_{2}$ & 0.999 & 0.98\tabularnewline
Adam, $\epsilon$ & 0.000001 & 0.000001\tabularnewline
\bottomrule
\end{tabular}
	\end{adjustbox}\VerticalSpaceFloat
\label{Tbl:hyperparam}
{\footnotesize
\textit{Notes:} This table shows the hyperparameters using to pre-train BERT and RoBERTa, respectively.\par}
\end{table} 
\FloatBarrier
\clearpage
\section{Degree of task measures}\label{App:tasks}
In this appendix, we provide additional details to those given in Section \ref{sec:Application} on the application. This appendix is meant as a supplement and is not self-contained without the main text. 

\begin{table}[!ht]
\caption{Components of the task measures}
	\begin{adjustbox}{max totalsize = {\textwidth}{0.9\textheight}, center}
\begin{tabular}{p{0.17\linewidth}p{0.26\linewidth}p{0.14\linewidth}p{0.7\linewidth}}
\toprule 
Scale & Subscale & Element ID & Element name\tabularnewline
\midrule
Abstract & Non-routine cognitive: Analytical & 4.A.2.a.4 & Analyzing data/information\tabularnewline
 &  & 4.A.2.b.2 & Thinking creatively\tabularnewline
 &  & 4.A.4.a.1 & Interpreting information for others\tabularnewline
 & Non-routine cognitive: Interpersonal & 4.A.4.a.4 & Establishing and maintaining personal relationships\tabularnewline
 &  & 4.A.4.b.4 & Guiding, directing and motivating subordinates\tabularnewline
 &  & 4.A.4.b.5 & Coaching/developing others\tabularnewline
 &  &  & \tabularnewline
Manual & Non-routine manual & 4.A.3.a.4 & Operating vehicles, mechanized devices, or equipment\tabularnewline
 &  & 4.C.2.d.1.g & Spend time using hands to handle, control or feel objects, tools or
controls\tabularnewline
 &  & 1.A.2.a.2 & Manual dexterity\tabularnewline
 &  & 1.A.1.f.1 & Spatial orientation\tabularnewline
 &  &  & \tabularnewline
Routine & Routine cognitive & 4.C.3.b.7 & Importance of repeating the same tasks\tabularnewline
 &  & 4.C.3.b.4 & Importance of being exact or accurate\tabularnewline
 &  & 4.C.3.b.8 & Structured v. Unstructured work (reverse)\tabularnewline
 & Routine manual & 4.C.3.d.3 & Pace determined by speed of equipment\tabularnewline
 &  & 4.A.3.a.3 & Controlling machines and processes\tabularnewline
 &  & 4.C.2.d.1.i & Spend time making repetitive motions\tabularnewline
\bottomrule
\end{tabular}
	\end{adjustbox}\VerticalSpaceFloat
\label{Tbl:task_measures}
{\footnotesize
\textit{Notes:} This table shows the task measures and their components.  Each measure is a standardized sum of its subscales, and equivalently, each subscale is a standardized sum of its individual elements.  See \cite{Acemoglu2011} for further details.\par}
\end{table}

\paragraph*{Abstract tasks}
\begin{definition}[Abstract tasks]\label{Def:abstract_taks}
\textquote{Abstract tasks are problem-solving and managerial tasks, which require intuition, persuasion, and creativity and are unstructured and nonroutine. These abstract tasks are performed by workers who have high levels of education and analytical capability, and they benefit from computers that facilitate the transmission, organization, and processing of information. Abstract tasks are characteristic of professional, managerial, technical, and creative occupations, such as law, medicine, science, engineering, marketing, and design.}
\end{definition}

\begin{table}[!ht]
\caption{Top and bottom 10 occupations on degree of abstract tasks}
	\begin{adjustbox}{max totalsize = {\textwidth}{0.9\textheight}, center}
		\begin{tabular}{lll}
\toprule
{} &                                             Top 10 &                                          Bottom 10 \\
Rank &                                                    &                                                    \\
\midrule
1    &                       Operations Research Analysts &                                  Bicycle Repairers \\
2    &                              Intelligence Analysts &                                 Rail Car Repairers \\
3    &                                  Graphic Designers &  Outdoor Power Equipment and Other Small Engine... \\
4    &                                     Mathematicians &                        Tire Repairers and Changers \\
5    &                                       Sociologists &                               Motorcycle Mechanics \\
6    &       Computer and Information Research Scientists &                Non-Destructive Testing Specialists \\
7    &  Fine Artists, Including Painters, Sculptors, a... &  Rail-Track Laying and Maintenance Equipment Op... \\
8    &                             Office Clerks, General &           Solar Thermal Installers and Technicians \\
9    &                       Search Marketing Strategists &                       Plasterers and Stucco Masons \\
10   &                        Word Processors and Typists &   Telecommunications Line Installers and Repairers \\
\bottomrule
\end{tabular}

	\end{adjustbox}\VerticalSpaceFloat
\label{Tbl:top_and_bottom10_abstract_task}
{\footnotesize
\textit{Notes:} This table shows the top and bottom 10 occupations on degree of abstract tasks estimated by the \OccToVecS framework as the cosine similarity between occupation embeddings and the embedding of abstract tasks.\par}
\end{table}

\begin{figure}[!ht]
\begin{adjustbox}{max totalsize = {\textwidth}{\SingleFigMaxHeight\textheight}, center}
\includegraphics[width = \textwidth]{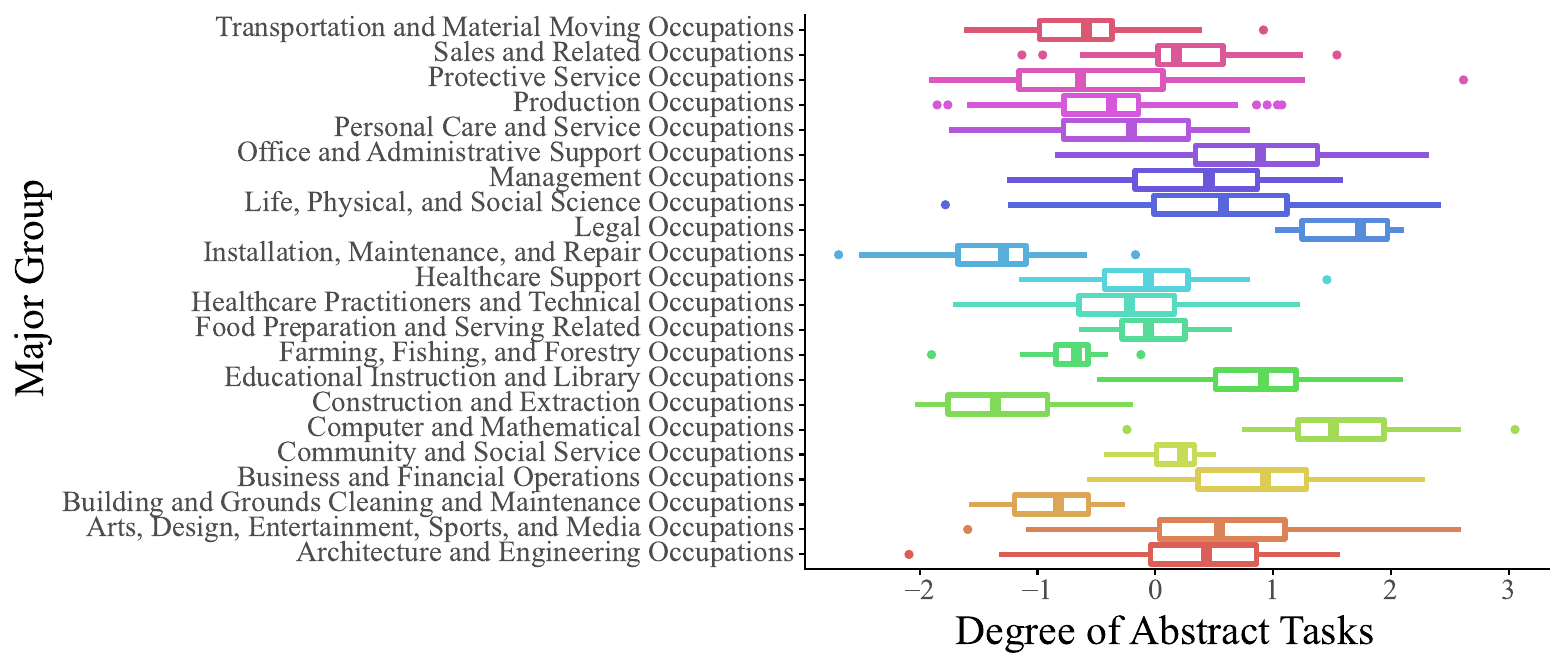}
\end{adjustbox}	
\caption{Abstract tasks by major occupational group}
\label{Fig:abstract_tasks_on_major_group_title}
\VerticalSpaceFloat
{\footnotesize
\textit{Notes:} This figure shows a boxplot of the occupational degree of abstract tasks by major occupational groups. \par}
\end{figure}
 
\begin{figure}[!ht]
\begin{adjustbox}{max totalsize = {\textwidth}{\SingleFigMaxHeight\textheight}, center}
\includegraphics[width = \textwidth]{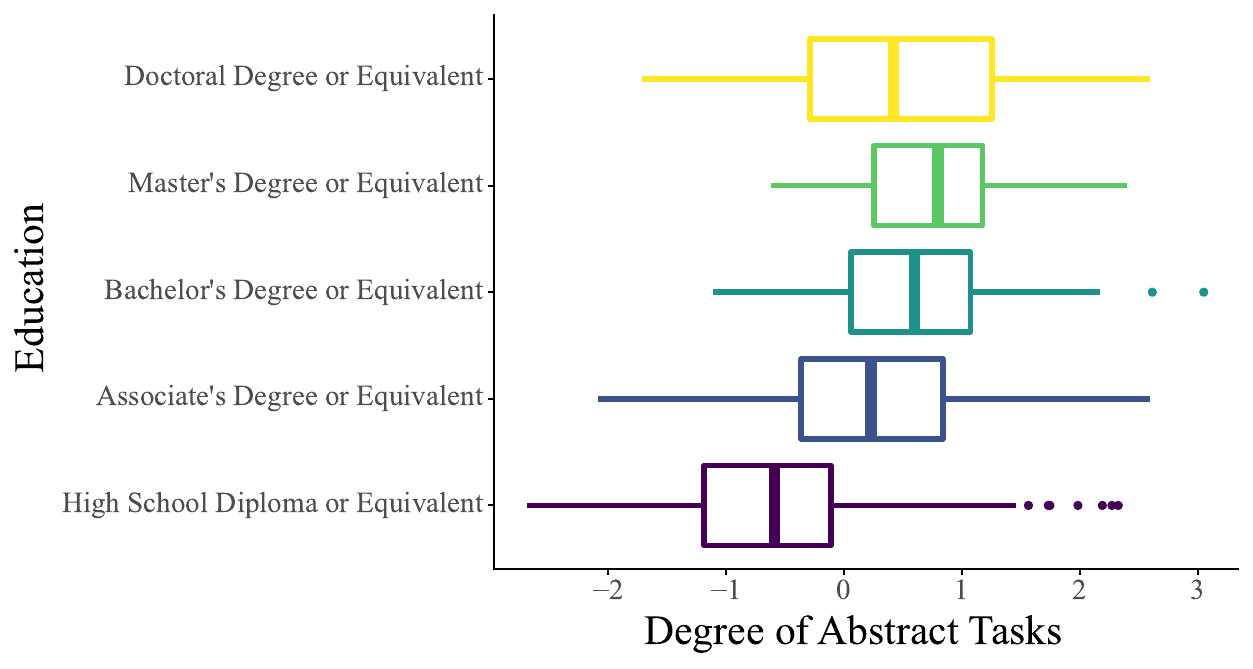}
\end{adjustbox}	
\caption{Abstract tasks by education}
\label{Fig:abstract_tasks_on_education}
\VerticalSpaceFloat
{\footnotesize
\textit{Notes:} This figure shows a boxplot of the occupational degree of abstract tasks by educational requirement. \par}
\end{figure}

\begin{figure}[!ht]
		\begin{adjustbox}{max totalsize = {\textwidth}{\SingleFigMaxHeight\textheight}, center}
			\includegraphics[width = \textwidth]{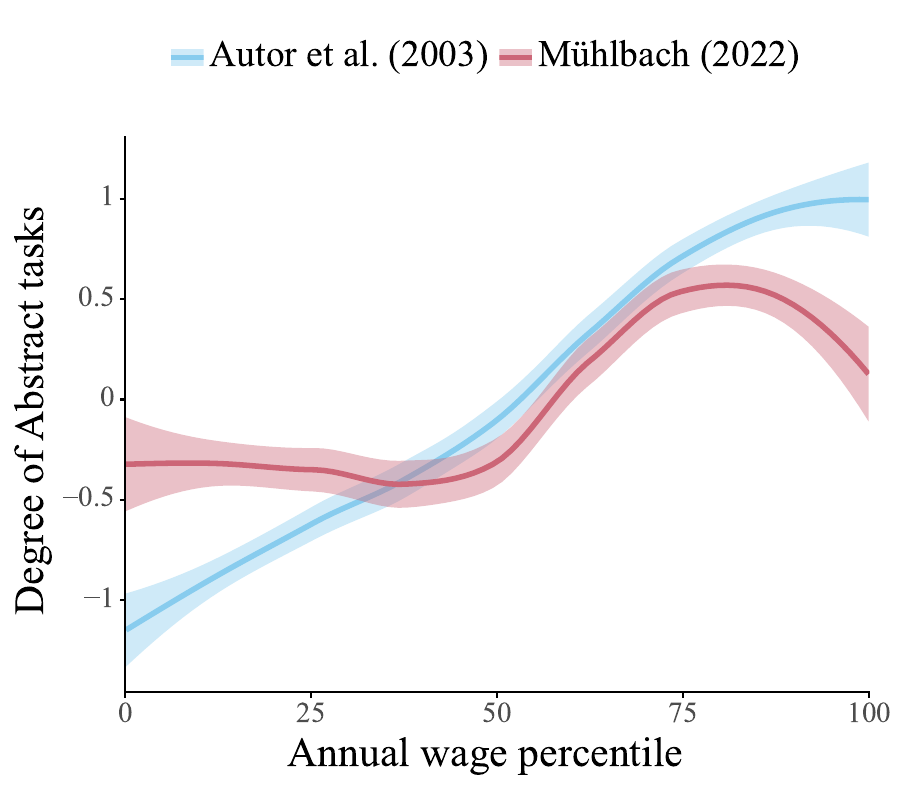}
		\end{adjustbox}	
	\caption{Abstract tasks by occupation wages}
	\label{Fig:abstract_tasks_on_annual_wage_percentile_by_methods}
\VerticalSpaceFloat
{\footnotesize
\textit{Notes:} This figure shows a smoothed polynomial regression of both standardized measures of abstract tasks in each 6-digit SOC occupation against its rank in the wage distribution.\par}
\end{figure}

\FloatBarrier
\paragraph*{Manual tasks}
\begin{definition}[Manual tasks]\label{Def:manual_taks}
\textquote{Manual tasks are innate tasks like dexterity, sightedness, and visual and language recognition, which require adaptability, flexibility, and in-person interactions. These tasks are typically nonpredictable but straightforward. Manual tasks are characteristic of construction and service occupations, such as truck drivers, janitors, and house-cleaners.}
\end{definition}

\begin{table}[!ht]
\caption{Top and bottom 10 occupations on degree of manual tasks}
	\begin{adjustbox}{max totalsize = {\textwidth}{0.9\textheight}, center}
		\begin{tabular}{lll}
\toprule
{} &                                             Top 10 &                                          Bottom 10 \\
Rank &                                                    &                                                    \\
\midrule
1    &  Laborers and Freight, Stock, and Material Move... &          Political Science Teachers, Postsecondary \\
2    &                                        Dishwashers &                      Chief Sustainability Officers \\
3    &             Industrial Truck and Tractor Operators &                    Financial Quantitative Analysts \\
4    &                              Helpers--Electricians &    Philosophy and Religion Teachers, Postsecondary \\
5    &  Cleaning, Washing, and Metal Pickling Equipmen... &                           Environmental Economists \\
6    &  Excavating and Loading Machine and Dragline Op... &       Mathematical Science Teachers, Postsecondary \\
7    &                    Maids and Housekeeping Cleaners &             Communications Teachers, Postsecondary \\
8    &                   Agricultural Equipment Operators &                        Law Teachers, Postsecondary \\
9    &                 Cleaners of Vehicles and Equipment &                     Climate Change Policy Analysts \\
10   &  Welding, Soldering, and Brazing Machine Setter... &  Agents and Business Managers of Artists, Perfo... \\
\bottomrule
\end{tabular}

	\end{adjustbox}\VerticalSpaceFloat
\label{Tbl:top_and_bottom10_manual_task}
{\footnotesize
\textit{Notes:} This table shows the top and bottom 10 occupations on degree of manual tasks estimated by the \OccToVecS framework as the cosine similarity between occupation embeddings and the embedding of manual tasks.\par}
\end{table}

\begin{figure}[!ht]
\begin{adjustbox}{max totalsize = {\textwidth}{\SingleFigMaxHeight\textheight}, center}
\includegraphics[width = \textwidth]{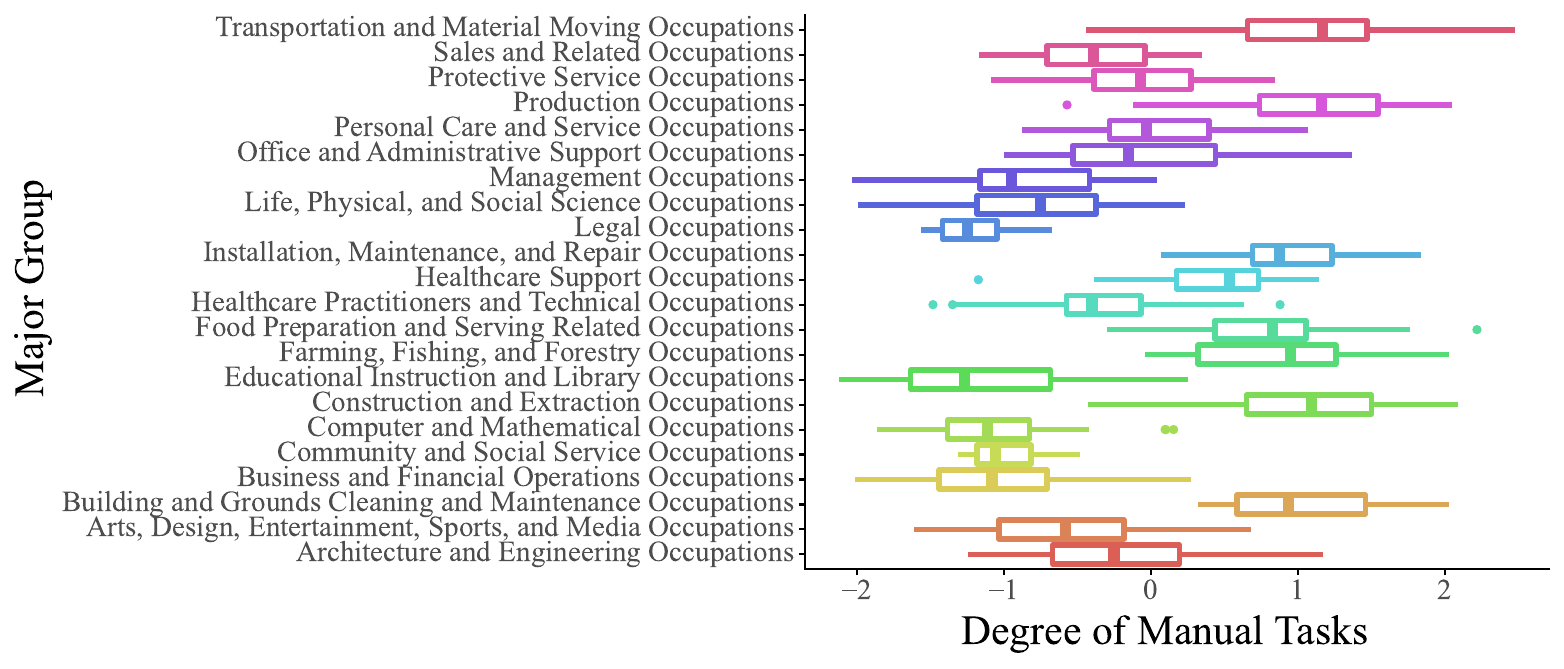}
\end{adjustbox}	
\caption{Manual tasks by major occupational group}
\label{Fig:manual_tasks_on_major_group_title}
\VerticalSpaceFloat
{\footnotesize
\textit{Notes:} This figure shows a boxplot of the occupational degree of manual tasks by major occupational groups. \par}
\end{figure}
 
\begin{figure}[!ht]
\begin{adjustbox}{max totalsize = {\textwidth}{\SingleFigMaxHeight\textheight}, center}
\includegraphics[width = \textwidth]{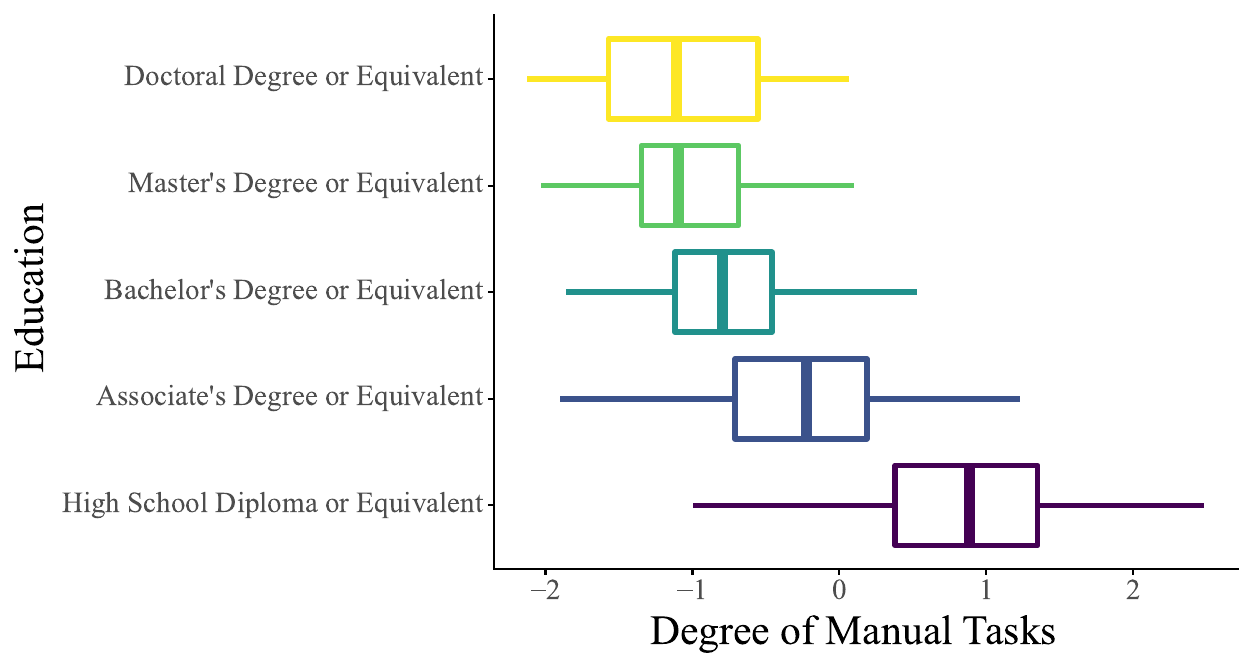}
\end{adjustbox}	
\caption{Manual tasks by education}
\label{Fig:manual_tasks_on_education}
\VerticalSpaceFloat
{\footnotesize
\textit{Notes:} This figure shows a boxplot of the occupational degree of manual tasks by educational requirement. \par}
\end{figure}

\begin{figure}[!ht]
		\begin{adjustbox}{max totalsize = {\textwidth}{\SingleFigMaxHeight\textheight}, center}
			\includegraphics[width = \textwidth]{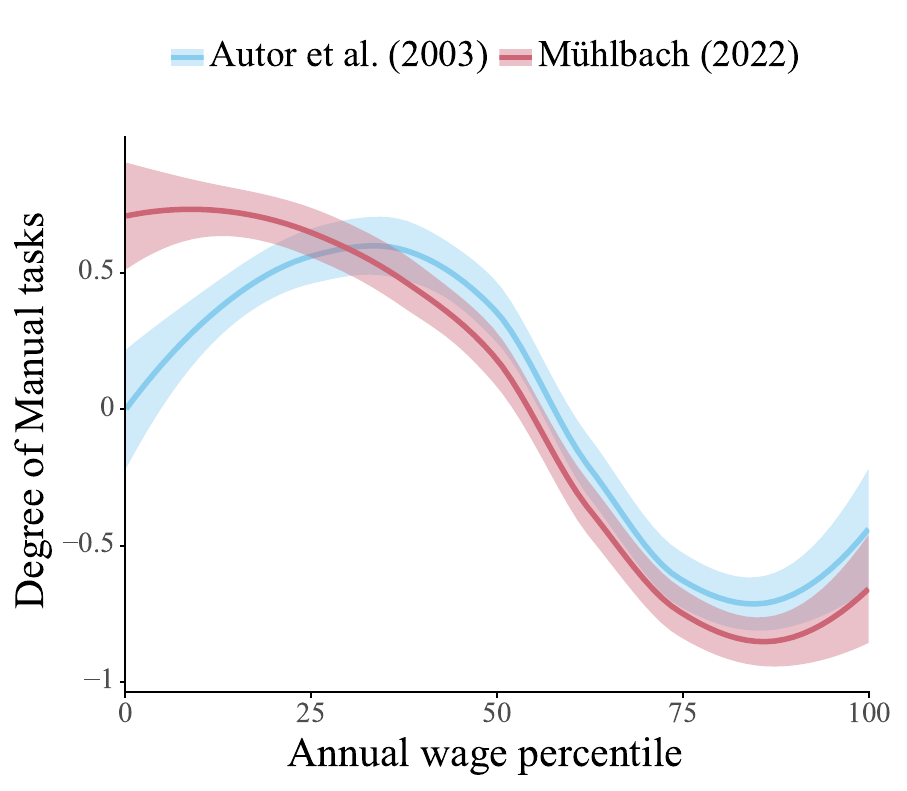}
		\end{adjustbox}	
	\caption{Manual tasks by occupation wages}
	\label{Fig:manual_tasks_on_annual_wage_percentile_by_methods}
\VerticalSpaceFloat
{\footnotesize
\textit{Notes:} This figure shows a smoothed polynomial regression of both standardized measures of manual tasks in each 6-digit SOC occupation against its rank in the wage distribution.\par}
\end{figure}

\FloatBarrier
\clearpage
\section{Degree of artificial intelligence}\label{App:ai}
In this appendix, we provide additional details to those given in Section \ref{sec:Application} on the application. This appendix is meant as a supplement and is not self-contained without the main text. 

\paragraph*{Definition of artificial intelligence}
The following definitions jointly define artificial intelligence.  The definitions are from IEEE Guide for Terms and Concepts in Intelligent Process Automation, IEEE Std 2755TM-2017 (\href{https://standards.ieee.org/standard/2755-2017.html}{link}). Starting from \textit{Artificial Intelligence} in Definition \ref{Def:artificial_intelligence}, we include the definitions of all related terms repeatedly until we have these nine definitions. By using our NLP algorithm, we get one embeddings for each definition, which we finally combine into one embedding by a simple average.

\begin{definition}[Artificial General Intelligence]\label{Def:artificial_general_intelligence}
\textquote{Complex, computational artificial intelligence capable of providing descriptive, discovery, predictive, prescriptive, and deductive analytics with relevance and accuracy equal to or exceeding human experts in multiple general knowledge domains. Artificial general intelligence includes artificial intelligence systems capable of interacting naturally with humans and machines in a way undetectable to expert observers and consistently passing the Turing Test for artificial intelligence.}
\end{definition}

\begin{definition}[Artificial Intelligence]\label{Def:artificial_intelligence}
\textquote{The combination of cognitive automation, machine learning, reasoning, hypothesis generation and analysis, natural language processing, and intentional algorithm mutation producing insights and analytics at or above human capability.}
\end{definition}

\begin{definition}[Machine Learning]\label{Def:machine_learning}
\textquote{The combination of cognitive automation, machine learning, reasoning, hypothesis generation and analysis, natural language processing, and intentional algorithm mutation producing insights and analytics at or above human capability.}
\end{definition}

\begin{definition}[Cognitive Automation]\label{Def:cognitive_automation}
\textquote{The identification, assessment, and application of available machine learning algorithms for the purpose of leveraging domain knowledge and reasoning to further automate the machine learning already present in a manner that may be thought of as cognitive. With cognitive automation, the system performs corrective actions driven by knowledge of the underlying analytics tool itself, iterates its own automation approaches and algorithms for more expansive or more thorough analysis, and is thereby able to fulfill its purpose. The automation of the cognitive process refines itself and dynamically generates novel hypotheses that it can likewise assess against its existing corpus and other information resources.}
\end{definition}

\begin{definition}[Descriptive Analytics]\label{Def:descriptive_analytics}
\textquote{Insights, reporting, and information answering the question, “Why did something happen?” Descriptive analytics determines information useful to understanding the cause(s) of an event(s).}
\end{definition}

\begin{definition}[Deductive Analytics]\label{Def:deductive_analytics}
\textquote{Insights, reporting, and information answering the question, “What would likely happen IF...?” Deductive analytics evaluates causes and outcomes of possible future events.}
\end{definition}

\begin{definition}[Diagnostic Analytics]\label{Def:diagnostic_analytics}
\textquote{Insights, reporting, and information answering the question, “Why did something happen?” Diagnostic analytics determines information useful to understanding the cause(s) of an event(s).}
\end{definition}

\begin{definition}[Predictive Analytics]\label{Def:predictive_analytics}
\textquote{Insights, reporting, and information answering the question, “What is likely to happen?” Predictive analytics support high confidence foretelling of future event(s).}
\end{definition}

\begin{definition}[Prescriptive Analytics]\label{Def:prescriptive_analytics}
\textquote{Insights, reporting, and information answering the question, “What should I do about it?” Prescriptive analytics determines information that provides high confidence actions necessary to recover from an event or fulfill a need.}
\end{definition}

\paragraph*{Top and bottom 10 occupations}
\begin{table}[!ht]
\caption{Top and bottom 10 occupations on degree of artificial intelligence}
	\begin{adjustbox}{max totalsize = {\textwidth}{0.9\textheight}, center}
		\begin{tabular}{lll}
\toprule
{} &                                             Top 10 &                                          Bottom 10 \\
Rank &                                                    &                                                    \\
\midrule
1    &                     Business Intelligence Analysts &  Helpers--Brickmasons, Blockmasons, Stonemasons... \\
2    &                                      Statisticians &  Helpers--Painters, Paperhangers, Plasterers, a... \\
3    &                              Intelligence Analysts &  Farmworkers and Laborers, Crop, Nursery, and G... \\
4    &                       Operations Research Analysts &                              Construction Laborers \\
5    &  Market Research Analysts and Marketing Special... &       Meat, Poultry, and Fish Cutters and Trimmers \\
6    &                    Financial Quantitative Analysts &                                             Tapers \\
7    &                       Data Warehousing Specialists &                               Roof Bolters, Mining \\
8    &                             Clinical Data Managers &                       Plasterers and Stucco Masons \\
9    &                                   Biostatisticians &                        Manicurists and Pedicurists \\
10   &                       Search Marketing Strategists &                                     Fence Erectors \\
\bottomrule
\end{tabular}

	\end{adjustbox}\VerticalSpaceFloat
\label{Tbl:top_and_bottom10_artificial_intelligence}
{\footnotesize
\textit{Notes:} This table shows the top and bottom 10 occupations on degree of artificial intelligence estimated by the \textbf{occupation2vec} framework.\par}
\end{table}

\paragraph*{Major occupational groups and educational requirements}

\begin{figure}[!ht]
\begin{adjustbox}{max totalsize = {\textwidth}{\SingleFigMaxHeight\textheight}, center}
\includegraphics[width = \textwidth]{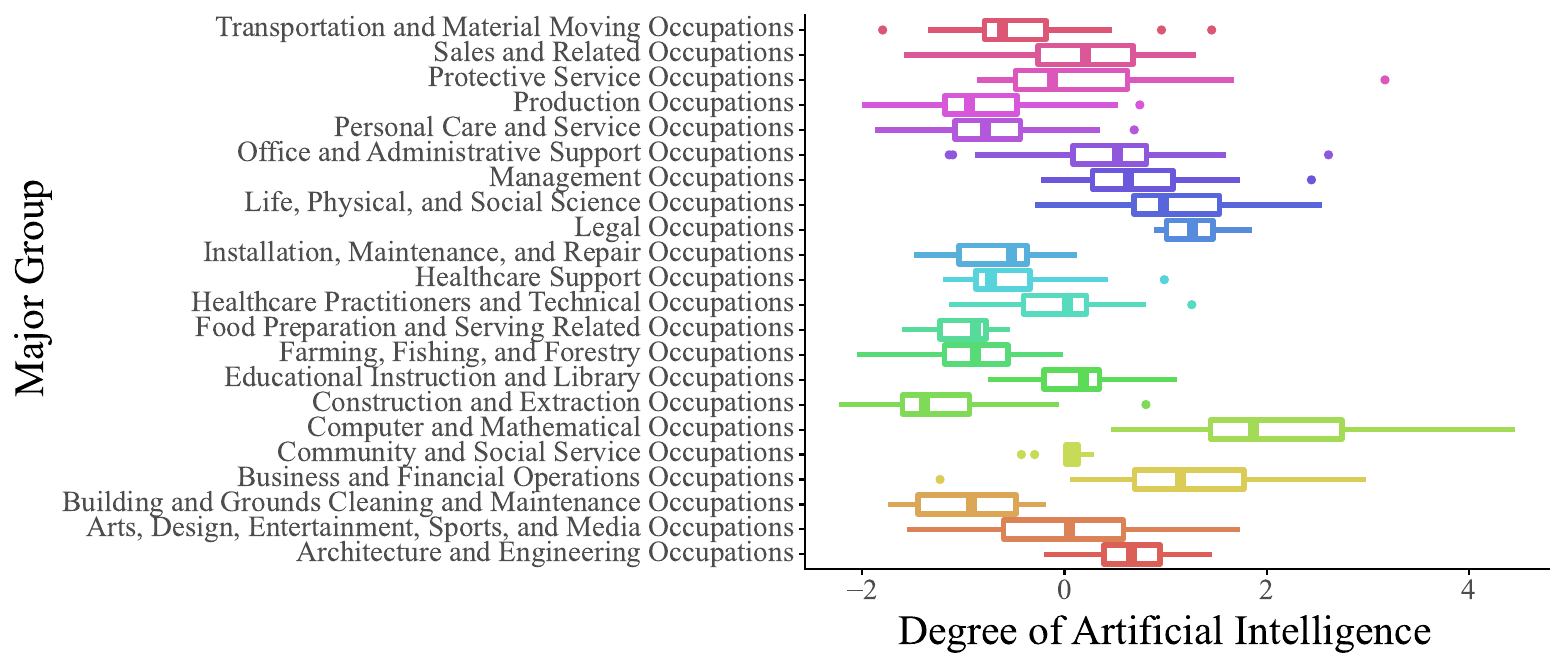}
\end{adjustbox}	
\caption{Artificial intelligence by major occupational group}
\label{Fig:artificial_intelligence_on_major_group_title}
\VerticalSpaceFloat
{\footnotesize
\textit{Notes:} This figure shows a boxplot of the occupational degree of artificial intelligence by major occupational groups. \par}
\end{figure}
 
\begin{figure}[!ht]
\begin{adjustbox}{max totalsize = {\textwidth}{\SingleFigMaxHeight\textheight}, center}
\includegraphics[width = \textwidth]{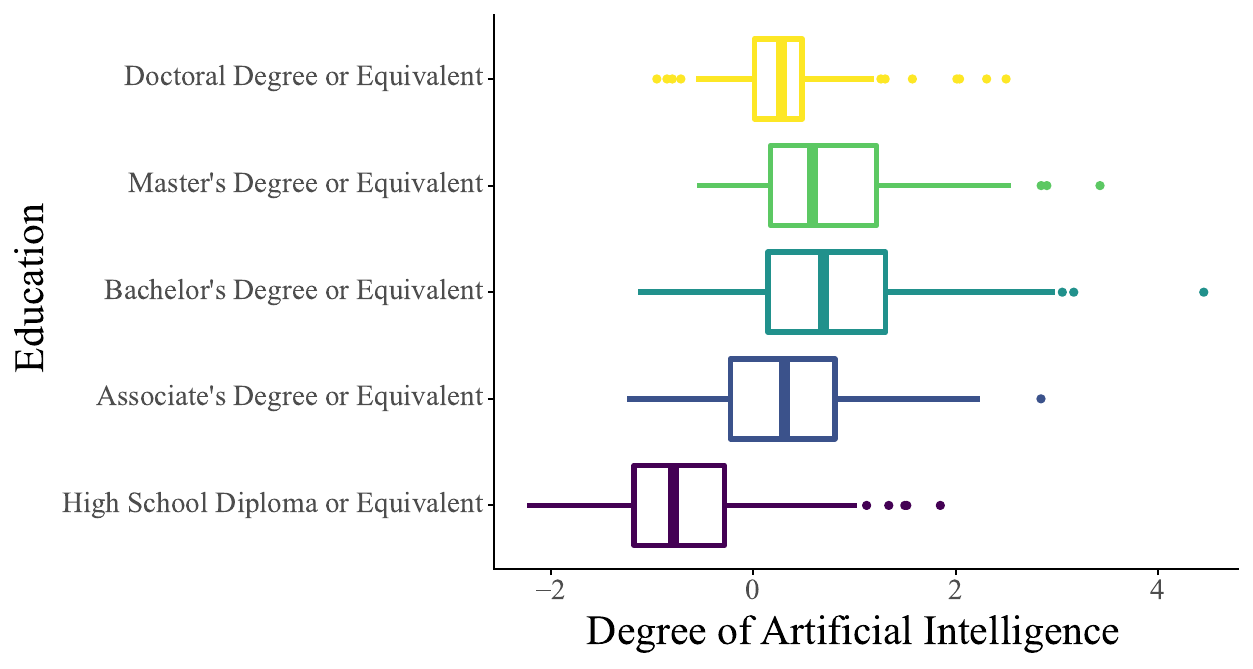}
\end{adjustbox}	
\caption{Artificial intelligence by education}
\label{Fig:artificial_intelligence_on_education}
\VerticalSpaceFloat
{\footnotesize
\textit{Notes:} This figure shows a boxplot of the occupational degree of artificial intelligence by educational requirement. \par}
\end{figure}

\paragraph*{Wage and employment growth profiles}

\begin{figure}[!ht]
		\begin{adjustbox}{max totalsize = {\textwidth}{\SingleFigMaxHeight\textheight}, center}
			\includegraphics[width = \textwidth]{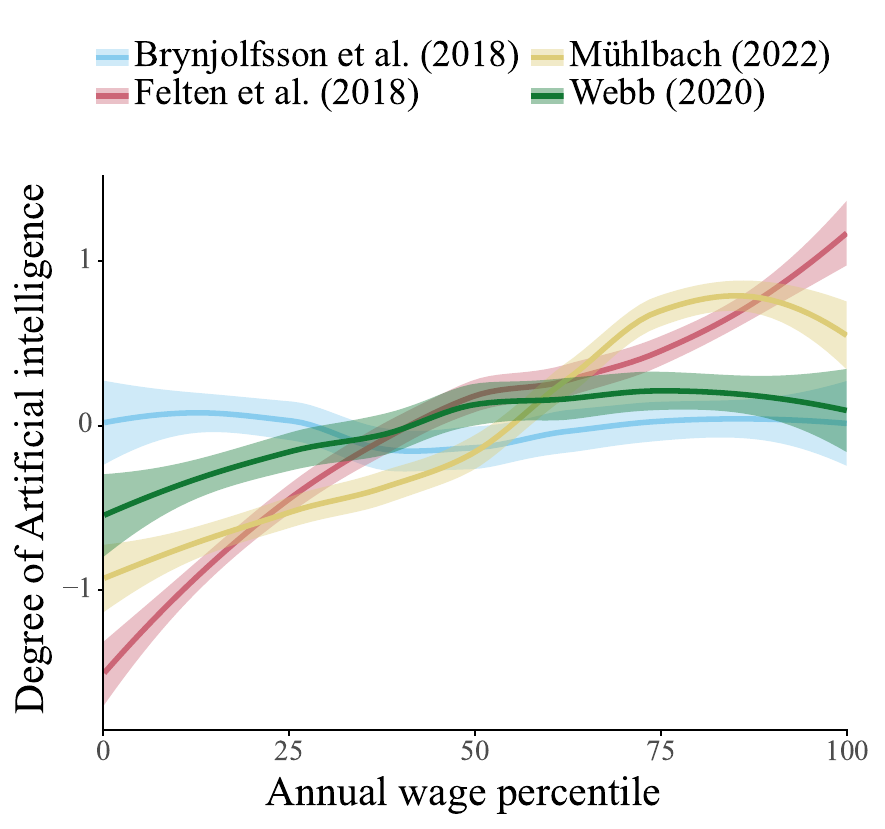}
		\end{adjustbox}	
	\caption{Artificial intelligence by occupation wages}
	\label{Fig:comparing_concepts_artificial_intelligence_on_stat_annual_wage}
\VerticalSpaceFloat
{\footnotesize
\textit{Notes:} This figure shows a smoothed polynomial regression of standardized measures of artificial intelligence in each 6-digit SOC occupation against its rank in the wage distribution.\par}
\end{figure}

\paragraph*{Correlation coefficients}

\begin{table}[!ht]
\caption{Correlation coefficients between measures}
	\begin{adjustbox}{max totalsize = {\textwidth}{0.9\textheight}, center}
		\begin{tabular}{lrrrr}
\toprule
{} &  Brynjolfsson et al. (2018) &  Felten et al. (2018) &  Mühlbach (2022) &  Webb (2020) \\
\midrule
Brynjolfsson et al. (2018) &                       1.000 &                -0.090 &            0.284 &        0.059 \\
Felten et al. (2018)       &                      -0.090 &                 1.000 &            0.353 &        0.391 \\
Mühlbach (2022)            &                       0.284 &                 0.353 &            1.000 &        0.202 \\
Webb (2020)                &                       0.059 &                 0.391 &            0.202 &        1.000 \\
\bottomrule
\end{tabular}

	\end{adjustbox}\VerticalSpaceFloat
\label{Tbl:correlation_coefficients_artificial_intelligence}
{\footnotesize
\textit{Notes:} This table shows the Spearman correlation coefficient from comparing our occupational estimates of artificial intelligence to the ones of \cite{Felten2018},  \cite{Brynjolfsson2018a},  and \cite{Webb2019}., respectively. \par}
\end{table}
 
\FloatBarrier
\clearpage
\section{Degree of charisma}\label{App:charisma}
In this appendix, we provide additional details to those given in Section \ref{sec:Application} on the application. This appendix is meant as a supplement and is not self-contained without the main text. 

\paragraph*{Definition of charisma}
The following definitions jointly define charisma.  The definitions are from Psychology Today,  Cambridge University Press, Oxford University Press, and Wikipedia, respectively.  In addition, we use American Psychological Association's Dictionary of Psychology (see Definition \ref{Def:charisma_apa}). By using our NLP algorithm, we get one embedding for each definition, which we finally combine into one embedding by a simple average.

\begin{definition}[Charisma]\label{Def:charisma_psyc}
\textquote{Charisma is an individual's ability to attract and influence other people. While it is often described as a mysterious quality that one either has or doesn't have, some experts argue that the skills of charismatic people can be learned and cultivated.} \citep{Psyc_charisma}
\end{definition}

\begin{definition}[Charisma]\label{Def:charisma_cambridge}
\textquote{A special power that some people have naturally that makes them able to influence other people and attract their attention and admiration.} \citep{Cambridge_charisma}
\end{definition}

\begin{definition}[Charisma]\label{Def:charisma_oxford}
\textquote{A quality possessed by some individuals that encourages others to listen and follow. Charismatic leaders tend to be self-confident, visionary, and change oriented, often with eccentric or unusual behavior.} \citep{Oxford_charisma}
\end{definition}

\begin{definition}[Charisma]\label{Def:charisma_wiki}
\textquote{Charisma is compelling attractiveness or charm that can inspire devotion in others. Scholars in sociology, political science, psychology, and management reserve the term for a type of leadership seen as extraordinary; in these fields, the term charisma is used to describe a particular type of leader who uses values-based, symbolic, and emotion-laden leader signaling.} \citep{Wiki_charisma}
\end{definition}
 
\FloatBarrier
\clearpage
\section{Degree of emotional intelligence}\label{App:eq}
In this appendix, we provide additional details to those given in Section \ref{sec:Application} on the application. This appendix is meant as a supplement and is not self-contained without the main text. 

\paragraph*{Definition of emotional intelligence}
The following definitions jointly define emotional intelligence.  The definitions are from American Psychological Association, Psychology Today,  Cambridge University Press, Oxford University Press, and Wikipedia, respectively.  By using our NLP algorithm, we get one embedding for each definition, which we finally combine into one embedding by a simple average.

\begin{definition}[Emotional intelligence]\label{Def:eq_apa}
\textquote{A type of intelligence that involves the ability to process emotional information and use it in reasoning and other cognitive activities, proposed by U.S. psychologists Peter Salovey and John D. Mayer. According to Mayer and Salovey's 1997 model, it comprises four abilities: to perceive and appraise emotions accurately; to access and evoke emotions when they facilitate cognition; to comprehend emotional language and make use of emotional information; and to regulate one's own and others' emotions to promote growth and well-being. Their ideas were popularized in a best-selling book by U.S. psychologist and science journalist Daniel J. Goleman who also altered the definition to include many personality variables.} \citep{APA_eq}
\end{definition}

\begin{definition}[Emotional intelligence]\label{Def:eq_psyc}
\textquote{Emotional intelligence refers to the ability to identify and manage one’s own emotions, as well as the emotions of others. Emotional intelligence is generally said to include a few skills: namely emotional awareness, or the ability to identify and name one’s own emotions; the ability to harness those emotions and apply them to tasks like thinking and problem solving; and the ability to manage emotions, which includes both regulating one’s own emotions when necessary and helping others to do the same.} \citep{Psyc_eq}
\end{definition}

\begin{definition}[Emotional intelligence]\label{Def:eq_cambridge}
\textquote{The ability to understand the way people feel and react and to use this skill to make good judgments and to avoid or solve problems.} \citep{Cambridge_eq}
\end{definition}

\begin{definition}[Emotional intelligence]\label{Def:eq_oxford}
\textquote{Ability to monitor one's own and other people's emotions, to discriminate between different emotions and label them appropriately, and to use emotional information to guide thinking and behavior. It encompasses four competencies: the ability to perceive, appraise, and express emotions accurately; the ability to access and evoke emotions when they facilitate cognition; the ability to comprehend emotional messages and to make use of emotional information; and the ability to regulate one's own emotions to promote growth and well-being.  ($\ldots$) Popularized interpretations of emotional intelligence include various other factors such as interpersonal skills and adaptability.} \citep{Oxford_eq}
\end{definition}

\begin{definition}[Emotional intelligence]\label{Def:eq_wiki}
\textquote{Emotional intelligence is most often defined as the ability to perceive, use, understand, manage, and handle emotions. People with high emotional intelligence can recognize their own emotions and those of others, use emotional information to guide thinking and behavior, discern between different feelings and label them appropriately, and adjust emotions to adapt to environments.} \citep{Wiki_eq}
\end{definition}

\paragraph*{Top and bottom 10 occupations}
\begin{table}[!ht]
\caption{Top and bottom 10 occupations on degree of emotional intelligence}
	\begin{adjustbox}{max totalsize = {\textwidth}{0.9\textheight}, center}
		\begin{tabular}{lll}
\toprule
{} &                                 Top 10 &                                          Bottom 10 \\
Rank &                                        &                                                    \\
\midrule
1    &         Marriage and Family Therapists &  Helpers--Brickmasons, Blockmasons, Stonemasons... \\
2    &  Clinical and Counseling Psychologists &                                            Riggers \\
3    &               Mental Health Counselors &  Helpers--Painters, Paperhangers, Plasterers, a... \\
4    &                   School Psychologists &                                             Tapers \\
5    &                 Insurance Underwriters &                     Maintenance Workers, Machinery \\
6    &               Investment Fund Managers &                                  Bicycle Repairers \\
7    &           Search Marketing Strategists &                 Reinforcing Iron and Rebar Workers \\
8    &           Business Continuity Planners &  Rolling Machine Setters, Operators, and Tender... \\
9    &         Business Intelligence Analysts &  Woodworking Machine Setters, Operators, and Te... \\
10   &                   Video Game Designers &  Rail-Track Laying and Maintenance Equipment Op... \\
\bottomrule
\end{tabular}

	\end{adjustbox}\VerticalSpaceFloat
\label{Tbl:top_and_bottom10_emotional_intelligence}
{\footnotesize
\textit{Notes:} This table shows the top and bottom 10 occupations on degree of emotional intelligence estimated by the \OccToVecS framework.\par}
\end{table}

\paragraph*{Major occupational groups and educational requirements}
\begin{figure}[!ht]
\begin{adjustbox}{max totalsize = {\textwidth}{\SingleFigMaxHeight\textheight}, center}
\includegraphics[width = \textwidth]{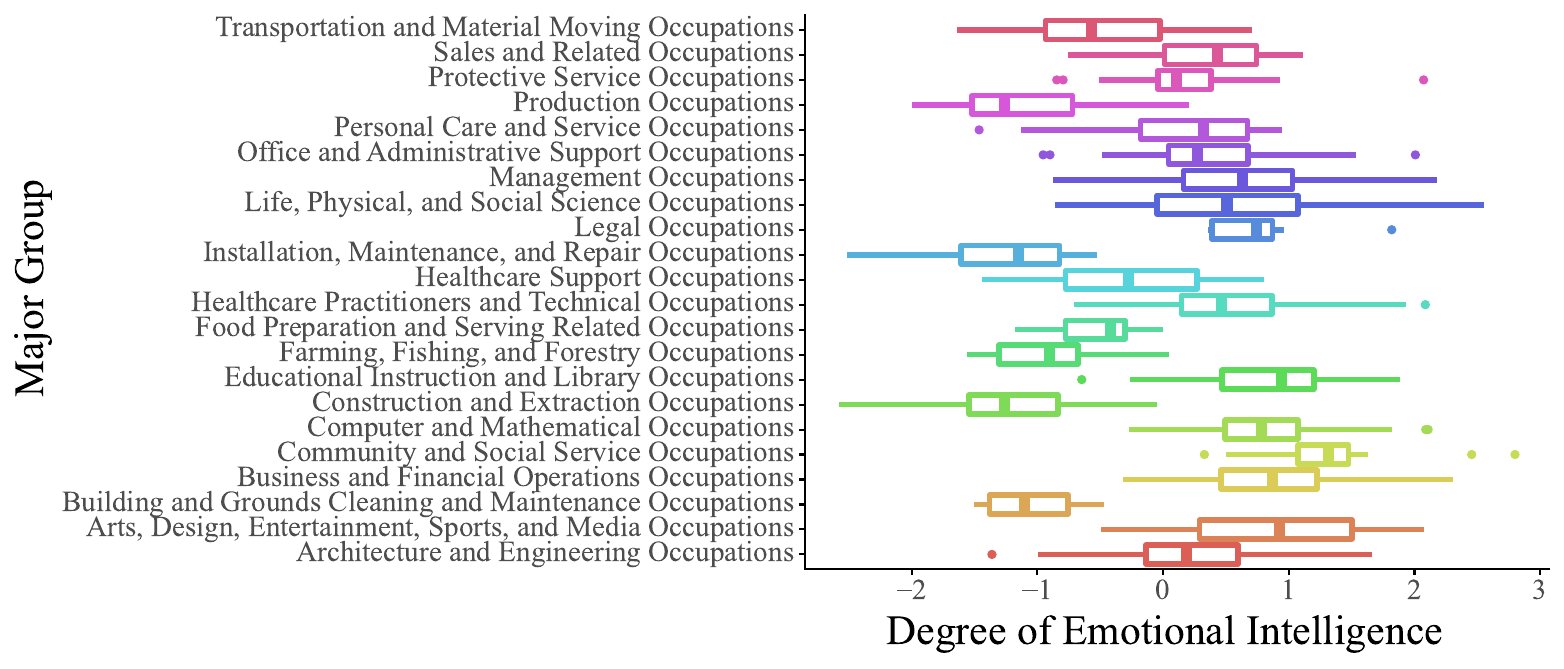}
\end{adjustbox}	
\caption{Emotional intelligence by major occupational group}
\label{Fig:emotional_intelligence_on_major_group_title}
\VerticalSpaceFloat
{\footnotesize
\textit{Notes:} This figure shows a boxplot of the occupational degree of emotional intelligence by major occupational groups. \par}
\end{figure}
 
\begin{figure}[!ht]
\begin{adjustbox}{max totalsize = {\textwidth}{\SingleFigMaxHeight\textheight}, center}
\includegraphics[width = \textwidth]{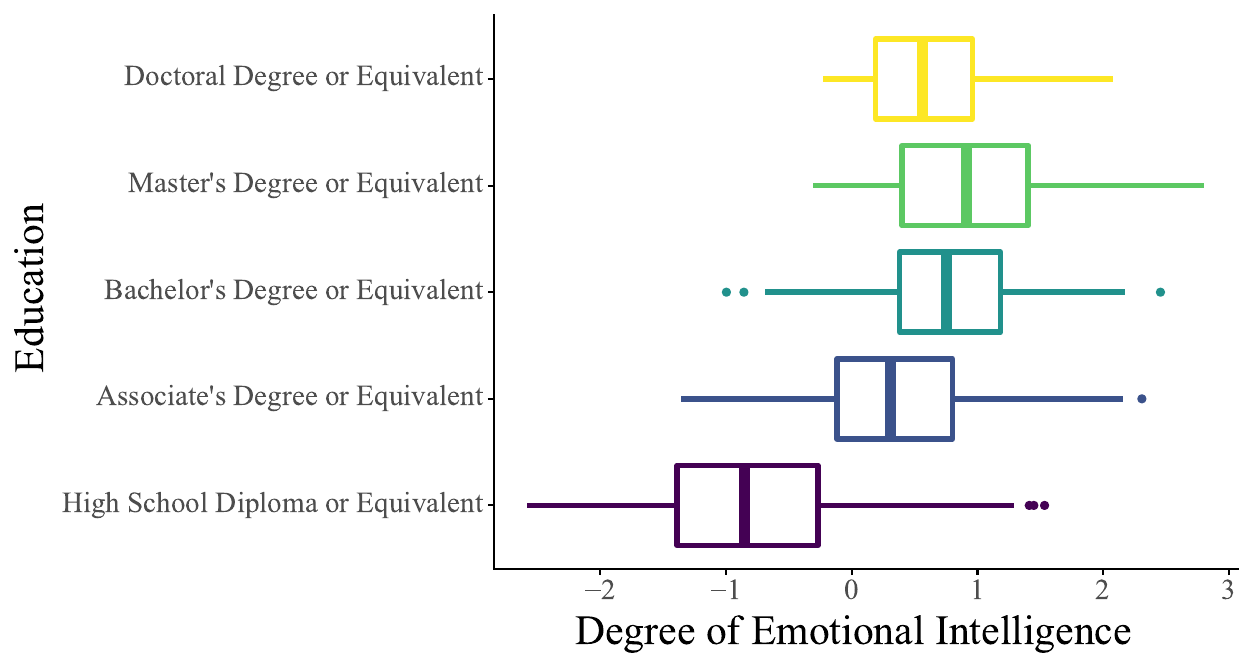}
\end{adjustbox}	
\caption{Emotional intelligence by education}
\label{Fig:emotional_intelligence_on_education}
\VerticalSpaceFloat
{\footnotesize
\textit{Notes:} This figure shows a boxplot of the occupational degree of emotional intelligence by educational requirement. \par}
\end{figure}

\paragraph*{Wage profiles}
\begin{figure}[!ht]
		\begin{adjustbox}{max totalsize = {\textwidth}{\SingleFigMaxHeight\textheight}, center}
			\includegraphics[width = \textwidth]{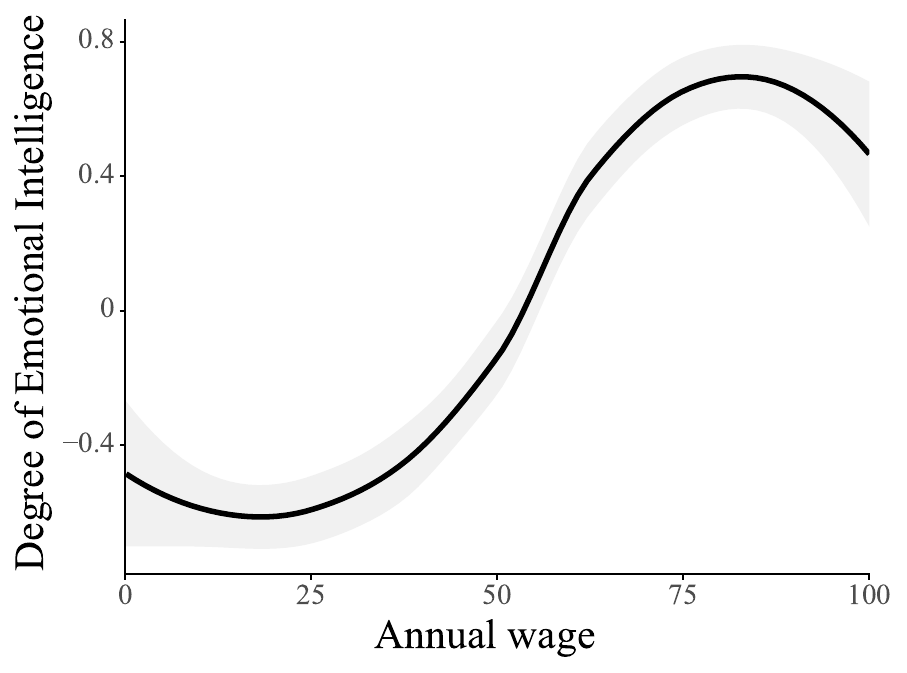}
		\end{adjustbox}	
	\caption{Emotional intelligence by occupation wages}
	\label{Fig:emotional_intelligence_on_annual_wage_percentile}
\VerticalSpaceFloat
{\footnotesize
\textit{Notes:} This figure shows a smoothed polynomial regression of our standardized measure of emotional intelligence in each 6-digit SOC occupation against its rank in the wage distribution.\par}
\end{figure}

\end{document}